\documentclass[fleqn,usenatbib]{mnras}
 \UseRawInputEncoding

\usepackage{amsfonts}
\usepackage{amsmath}
\usepackage{amssymb}
\usepackage{multirow}
\usepackage{ulem}
\usepackage{cancel}
\usepackage{graphicx}
\usepackage{color}
\usepackage[T1]{fontenc}

\DeclareRobustCommand{\VAN}[3]{#2}
\let\VANthebibliography\thebibliography
\def\thebibliography{\DeclareRobustCommand{\VAN}[3]{##3}\VANthebibliography}

\newcommand{\be}{\begin{equation}}
\newcommand{\ee}{\end{equation}}
\def\lsim{\mathrel{\rlap{\lower4pt\hbox{\hskip0.5pt$\sim$}}
    \raise1pt\hbox{$<$}}}         
\def\gsim{\mathrel{\rlap{\lower4pt\hbox{\hskip0.5pt$\sim$}}
    \raise1pt\hbox{$>$}}}         

\def\dd{{\rm d}}

\def\lsim{~\rlap{$<$}{\lower 1.0ex\hbox{$\sim$}}}
\def\bsim{~\rlap{$>$}{\lower 1.0ex\hbox{$\sim$}}}
\def\kms{\ {\rm km\,s^{-1}}}

\def\kpc{\ {\rm Kpc}}

\def\apjs{{Astrophys.~ J.~ Suppl.~}}
\def\apjl{{Astrophys.~ J.~ Lett.~}}

\def\cP{{\cal P}}
\def\fg{f_\textrm{g}}

\def\cL{\mathcal{L}}
\def\cGgr{\mathcal{G}_\textrm{gr}}
\def\cGsn{\mathcal{G}_\textrm{SN}}

\def\st{\sigma_{t}}
\def\Sg{\Sigma_{g}}
\def\sg{\sigma_{g}}

\def\Sfr{\Sigma_{SFR}}
\def\Sast{\Sigma_{\ast}}
\def\sast{\sigma_{\ast}}
\def\St{\widetilde \Sigma}

\def\kms{\, {\rm km }\, {\rm s}^{-1}}
\def\kpc{\, {\rm kpc }}

\def\Sigunits{\, {\rm M}_\odot\, {\rm pc}^{-2}}
\def\epsf{\epsilon_{_{SF} } }
\def\tsf{t_{_{SF}}}
\def\vsn{V_{SN}}
\def\etat{\eta}
\def\rhog{\rho_{g}}
 \def\sigQ{\sigma_{_{\!Q=1}}}
 \def\vh{v_\textrm{h}}
 \def\sigth{\sigma_{th}}
 \def\sigt{\sigma_{t}}
 \def\Sgth{\, \Sg^\textrm{thr}}

 \def\etavQ{4.5}
  \def\etavH{6.7}
  \def\epsvQ{0.007}
  \def\epsvH{0.013}
\definecolor{RedWine}{rgb}{0.743,0,0}
\definecolor{RoyalBlue}{rgb}{0.25,.41,.88}
\definecolor{ForestGreen}{rgb}{.13,.54,.13}
\definecolor{DeepPurple}{rgb}{.72,.18,1}

\title[Star formation and {gravito-turbulence} ]{Regulation of  star formation by large scale {gravito-turbulence} }

\author[Nusser \& Silk]{Adi Nusser$^{1}$ \  and Joseph Silk$^{2,3,4}$   \\
$^{1}$ 
Department of Physics and the Asher Space Research Institute, Israel Institute of Technology Technion, Haifa 32000, Israel \\
$^{2}$ Institut d'Astrophysique de Paris, Sorbonne Universit\'es, UPMC Univ. Paris 06 et CNRS, UMR 7095, F-75014, Paris, France\\
$^{3}$ Department of  Physics \& Astronomy, The Johns Hopkins University, Baltimore, MD 21218, USA\\
$^{4}$ Beecroft Institute of Particle Astrophysics and Cosmology, Department of Physics,
University of Oxford,  Oxford OX1 3RH, UK }

\begin{document}
\label{firstpage}
\pagerange{\pageref{firstpage}--\pageref{lastpage}}
\maketitle

\begin{abstract}
A simple model for star formation based on supernova (SN) feedback and gravitational heating via the collapse of  perturbations in gravitationally unstable disks reproduces
the Schmidt-Kennicutt relation between  the star formation rate (SFR) per unit area, $\Sfr$, and the gas surface density, $\Sg$, remarkably well. The gas velocity dispersion, $\sg$, is derived self-consistently in conjunction with  $\Sfr$ and is  found to match the observations. Gravitational  instability  triggers {``gravito-turbulence"}  at the scale  of the least stable perturbation mode, 
boosting  $\sg$ at 
 $\Sg\gsim \Sgth =50\Sigunits$, and contributing to the pressure needed to carry the disk weight vertically.  $\Sfr$ is reduced to the observed level at $ \Sg \gsim \Sgth$, whereas
 at lower surface densities,  SN feedback is  the prevailing  energy  source. 
Our proposed star formation recipes  require  efficiencies of order 1\%, and 
the Toomre parameter, $Q$, for the joint gaseous and stellar disk 
is predicted to be close to the critical value for marginal stability for $\Sg\lsim \Sgth$, spreading to lower values and larger gas velocity dispersion at  
higher $\Sg$.
\end{abstract}
\begin{keywords}
{Galaxies: star formation-- galaxy formation-- ISM  }
\end{keywords}

\section{introduction}

Despite the complex interplay between a multitude of physical processes involved in galaxy formation, 
 evolved galaxies exhibit remarkably simple  scaling relations. 
 One of these relations is  the Schmidt-Kennicutt star formation law in disk galaxies 
$\Sfr=A\Sg^n$ \citep{Schmidt1959,Kennicutt1998}  between the SFR per unit disk area, $\Sfr$, and the gas surface density, $\Sg$. The universality of this relation over many decades in $\Sg,$ and its low scatter, indicate that it might be a consequence of  a global quasi-equilibrium maintained over a long term evolution of galaxies.

If supernova (SN) feedback is the main driver for pressure, $p$, in the disk,
then gravity-pressure balance in the vertical direction to  the disk 
provides an appealing  argument to derive a relation between $\Sfr$ and $\Sg$.
However, if $p\propto  \Sfr$, vertical balance 
yields a near quadratic dependence  on gas surface density, deviating strongly, especially at  high $\Sg $, from the $n\sim 3/2$ power inferred from the observed SK law. 
Further, a conceptual issue is that 
the pressure balance argument is phenomenologically persuasive but is  insufficient to account
 both for the SFR and the  turbulence velocity dispersion  that supports the gas  disk. Self-regulation of star formation and SN momentum injection   are  locally incomplete   until interstellar turbulence is regulated.  Specifically, pressure balance relies on, but does not specify, the turbulent gas velocity dispersion. 
 
 One problem is that there is no feedback loop between star formation  and turbulent velocity dispersion in the simplest theory. 
 SN-momentum regulation  of star formation  cannot self-consistently regulate the turbulence in the absence of dissipation physics. Moreover, observationally, there is indeed evidence that local star formation efficiency  decreases with increasing turbulence  inside molecular clouds (MCs) \citep{Leroy2017,Querejeta2019}.

 In the absence of  feedback, SFRs are well known to be far too high, both in disk galaxies as well as in spheroidal systems.
Supernova-driven turbulence is the favored  mechanism that  
regulates star formation \citep{Joung2006}.

 Massive stellar feedback  equally plays a role. Both OB winds and SN are  essential in star--forming MCs to obtain the SFE required for the global SK law \citep{Hopkins2011}.
More recent cosmological simulations of disk galaxies have improved resolution and  focus on feedback via turbulent pressure in the inhomogeneous interstellar medium (ISM) \citep{Gurvich2020}.

Supernova-driven  momentum deposition is the leading candidate for feedback in cosmological simulations of star-forming galaxies,  but recovering the global SK relation and in particular its low observed scatter remains a challenge. 
Recent disk star formation simulations  are consistent with vertical balance of 
 disk weight and ISM pressure generated by momentum input from SN-regulated star formation, although only a quasilinear SK law with large scatter is produced. This is  consistent with resolved  galaxy studies on subkpc scales 
\citep{Brucy2020}
 but does not account for the global  SK law \citep{Kim2020}.
 
SN-driven turbulence is inevitably significant
in  the densest star-forming  regions such as the Central Molecular Zone of the MWG  and more generally  in starbursts \citep{Seifried2018}.
This  
is less effective in forming ellipticals or in dwarfs. 
Nevertheless,  in gas-rich systems, additional turbulent input seems to be  required. One possibility is dynamically-induced turbulence  \citep[e.g.][]{Fuchs1998,Semelin2000,Brucy2020},  e.g. by  external shear. Generally, SFRs in simulated disks with $\Sg\lsim 100 \Sigunits $ can be explained by SN feedback alone, but mass loading and disk outflows remain a problem, suggesting that additional feedback is required \citep{Smith2018}.
Incorporating pre-supernova massive star feedback enhances  the energetic impact of SNe on the ISM \citep{Lucas2020} and helps in producing outflows. But the presence of outlows does not seem to affect the SK inferred law as the SFR tends to adapt to the varying gas content (see Fig.~8 in \cite{Smith2018}).
Multiphase modelling further boosts the outflow rates \citep{Kim2020a}.

 Self-gravity, halo gravity, magnetic fields, radiation, stellar winds,  cosmic rays, and 
SN feedback all contribute to the energy stored in galactic disks. 
The interplay between these factors dictates the stability of the gaseous 
disk component and its susceptibility to form stars.
\cite{Krumholz2016} and \cite{Krumholz2018} consider a model where gravito-turbulence is 
sustained by gravitational energy released via  inflow of disk gas mass.
Inward mass transport  is detected in the outer parts of the disks of some galaxies in the HI Nearby Galaxy Survey \citep{Schmidt2016}. 
This model exploits  vertical gravity-pressure equilibrium and  energy balance between 
energy injection and dissipation.
Turbulence driven by both  gas transport and  SN feedback is needed to account for 
observations of the SFR and velocity dispersion \citep[see also][]{Yu2021}.

Even in the absence of star formation, the growth and collapse of  perturbations in   a gravitationally  unstable disk  leads to  a boost in the velocity dispersion of the gas and stars,
 pushing the disk towards   a state of marginal stability   \citep[e.g.][]{Lin1987,Bertin2001,Booth2019,Bethune2021}. 
{Starting with small perturbations of an otherwise homogeneous  cool gas disk, non-linear evolution leads to bulk motions and the appearance of  shocks heating the gas. However, radiative cooling of  shocks with temperatures well  above $ 10^3-10^4 \; \textrm{K}$, will  
leave  the gas with a low  thermal velocity dispersion. 
The interaction of 
large-scale shocks can also produce vortical flows generating  solenoidal turbulence cascading to small scales \citep{Elmegreen2003,Bournaud2010,Combes2012,Nandakumar2020}. 
In a clumpy disk made of gas clouds/clumps of sizes smaller that of the 
fastest growing perturbation, the dynamics can be different.
In this case, part of the energy in bulk motions is transformed to the velocity dispersion of gas clouds, via cloud-cloud interactions occurring at the crossings of 
 large-scale perturbations.
In effect, bulk motions  on the scale of gravitational instabilities  lead to turbulent motions  cascading 
 down to small scales likely dominated by gas clouds/clumps . 
 } We will use the generic term {gravito-turbulence}  to refer  to gravity driven turbulence.

There is compelling evidence for the importance of gravitational instabilities in 
governing the complex structure of neutral hydrogen in cases such as the disk galaxy IC 342 \citep{Crosthwaite2000}.
We develop
simple recipes for turbulence pressure driven  by gravitational instabilities  in galactic disks are found to reduce the SFR needed to establish vertical balance between disk weight and pressure, while  
the velocity dispersion associated with this {gravito-turbulence}  is  elevated 
above the prediction from SN-feedback alone to a level comparable to that inferred from  the observations.

The outline of our  paper is as follows. We review energy balance in the gas disk before discussing  gravitational instability in a vertically balanced disk of gas and stars. Two alternative star formation recipes are set out, and we study energy balance  and turbulence generation for disk driving by combined gravitational instability and supernova feedback. The velocity dispersion is predicted for low and high SFRs in disks both above and below a threshold gas surface density that is set by supernova regulation.
We self-consistently recover the observed star formation law.
For convenience we provide a summary of our notation in  table \ref{tab:gloss}.
\begin{table*}
\centering
\caption{Summary of variables}
\begin{tabular}{ |l|l|c|l| } 
\hline
notation & & units  & relevance/Eqs \\ 
\hline 
 $\Sg$ & gas surface density & $\Sigunits$ & disk stability, height, SFR \\ 
 &&&\\
$\Sast$ & stellar surface density & $\Sigunits$ & disk stability, height, SFR \\ 
 &&&\\
 $\Sfr$ & star formation rate per unit disk area & $\Sigunits \, {\rm Myr}^{-1}$ & predicted by model \\ 
 &  & ${\rm M}_\odot\, {\rm kpc}^{-2} \, {\rm yr}^{-1}$ & SFR law, Eq.~\ref{eq:sfrtsf} \\ 
 &&&\\
 $\Sgth$ & $\sim 50 \Sigunits$, threshold $\Sg$ below which   & $\Sigunits $ & predicted by model \\ 
  & heating is dominated by SN feedback &  &  \\ 
  &&&\\
  $\rho_g$ &  gas volume density  & $ {\rm M}_\odot \, {\rm pc}^{-3}$ &  -\\
  &&&\\
 $\sg$ & total (thermal+turbulent) gas velocity dispersion & $\kms$ &  disk stability, height, SFR\\
 &&&\\
 $\st$ &  turbulent gas velocity dispersion & $\kms$ &  disk stability, height, SFR\\
  &&&\\
 $\sast$ & stellar velocity dispersion & $\kms$ & disk stability, height, SFR \\
  &&&\\
  $c_s$ & effective velocity dispersion for & $\kms$  & growth rate and scale of perturb.  \\
 &  gas-star composite disk  &  & proposed approx. in Eq.~\ref{eq:ceff} \\
  &&&\\
 $Q_g $ & Toomre stability parameter for gas disk only & - & Eq.~\ref{eq:Qgas} \\
  &&&\\
 $Q_s$ & Toomre stability parameter for stellar disk only & - & Eq.~\ref{eq:Qgas}\\
  &&&\\
  $Q$ & Toomre stability parameter for  & - & approx. Eq.~\ref{eq:Qeff} by \cite{Romeo2011} \\
 & gas-star composite disk  &  & \\
  &&&\\
  $H_0$ &  gas disk height (halo gravity not included) & kpc & Eq.\ref{eq:dHnoh} \\
    &&&\\
 $H$ &  gas disk height (halo gravity included) & kpc & turbulence driving scale in $\rho-$recipe \\
 &  &  & new approx. Eq.~\ref{eq:dHwh} \\
   &&&\\
$k_1^{-1}$ & physical scale of least stable perturb. in & kpc & turbulence driving scale in $Q-$recipe\\
 & composite disk & & new approx. Eq.~\ref{eq:kone} \\
&&&\\
$\omega_1^{-1}$ & timescale of  least stable mode in & ${\rm Myr}$ & turbulence driving scale in $Q-$recipe\\
 & composite disk & & new approx. Eq.~\ref{eq:tgrow} \\
&&&\\
$\tsf$ & timescale for gas collapse into stars & Myr & $ {G \rho_g}^{-1/2}$ in $\rho-$recipe, $  \omega_1^{-1}$ in  $Q-$recipe \\
 &  &  &  Eqs.~\ref{eq:sfrtsf}, \ref{eq:tsfQ} \& \ref{eq:tsfrho}\\
 &&&\\
 $\epsf$ & star formation efficiency & - & fraction of gas that de facto  turns  \\
 &  & &  into stars over timescale $\tsf$ \\
 &&&\\
 $\etat$ & turbulence dissipation timescale & - & Eq.~\ref{eq:tloss}  \\
\hline
\end{tabular}
\label{tab:gloss}
\end{table*}

\section{Energy balance}
\label{sec:Ebalance}
Consider the energy balance in the gas disk between supernova remnant heating, turbulent heating and gas thermal cooling. The energy balance equation  is
\begin{equation}
\label{eq:Eschematic}
\frac{3}{2}\frac{\dd \Sg \sg^2} {\dd t}=\left(\cGsn+\cGgr-\cL \right)\Sg\; . 
\end{equation}
The l.h.s represents the change in the kinetic  energy of the disk per unit area, where $\sg$ is the 1D gas velocity dispersion.
We write 
\begin{equation}
\sg^2=\sigth^2+\sigt^2\; , 
\end{equation}
where $\sigth$ is the contribution from thermal motions and $\sigt$ is due to turbulence. 
It is assumed that a thermal floor exists as a result of inefficient radiative cooling below a certain temperature in a mixture of neutral and molecular hydrogen \citep{Krumholz2018}. The precise temperature threshold depends on the detailed chemical composition of the gas, varying from $\sim 10$, $ \sim 100$, $\sim 1000$ K in $z\sim 0$ molecular, atomic and $z\sim 20$ primordial clouds, respectively. Thus $\sigth $ ranges from 
$\sigth \sim 0.2\kms$ for molecular gas to $\sim 5 \kms $ for atomic hydrogen \citep{Krumholz2018}. 
Turbulent motions generated on large scales cascade down until they dissipate to thermal energy on small scales. 
  The  term $\cGsn$ is due to SN  energy input. In principle, this term should include thermal energy as well as momentum injected into the ISM.   Most of the thermal energy is radiated away before the SN remnant merges with the ISM so that  momentum injection is dominant \citep{Cioffi1988}. However, it is important to note that in the case of a significant overlap of SN remnants (a large SN porosity or clustered SN resulting in superbubbles), hot high entropy ISM gas can produce galactic fountains and outflows \citep{MacLow1988,Keller2014,Kim2016}. 
The term $ \cGgr$ represents the  boost in the kinetic energy of collapsing perturbations. The 
  dissipation term $\cL$ refers mainly to loss of turbulent energy in compressible turbulence \citep{Federrath2009}.

This equation assumes that  there is no global change in the potential energy of the disk as a result of SF and 
local gravitational collapse. 
Nor are other energy components included, e.g. cosmic rays \citep{Wentzel1971} and magnetic fields  \citep[c.f.][for a review]{Beck2016}. 
 
 In order to derive specific forms for the energy gain and loss terms, we first present analyses of  disk instability and vertical gravity-pressure balance.

\section{Disk instability}

Gravitational instability of perturbations in the disk plays
an important role in the self-regulation of star formation.  
A  stable disk does not form stars efficiently    and a highly unstable disk  results in excessive star formation,
in turn self-regulating by raising the disk pressure towards  marginal stability.
Galactic disks can develop several types of instabilities related to the gas, magnetic field and collisionless components \citep[e.g.][]{Goldreich1965,Lynden-Bell1966,Julian1966,Lovelace1978,Elmegreen1983,Balbus1985,Balbus1998,Umurhan2010,Lovelace2013}.
Here we focus on the gravitational  instabilities of  axisymmetric perturbations in rotating disks \cite{Toomre1964}.

In a thin single fluid disk made of gas, the stability of linear 
perturbations  depends on the
 Toomre parameter 
 \begin{equation}
 \label{eq:Qgas}
 Q_{g} =\frac{\kappa \sg}{\pi G \Sg}\; , 
 \end{equation}
 where  $\kappa $ is the epicyclic frequency   given by $\kappa^2 =(2\Omega/R)d(R^2\Omega)/d \Omega$  with $R$ the radius inside the plane of the disk and $\Omega$ is the angular speed of circular orbits at $R$. 
We will adopt $\kappa=\sqrt{2} \Omega$ appropriate for a flat rotation curve.
 The perturbation is stable if
$Q_g\ge 1$  and unstable otherwise.
Although under  certain conditions, non-axisymmetric perturbations can grow even for $Q_{g} $ of the order of a few
\citep[e.g.][]{Goldreich1965,Lovelace1978,Lovelace2013},  generally instabilities associated 
with low $Q_{g}$ are usually dominant \citep[but see][]{Inoue2016}.

Realistic disks are multicomponent containing  warm gas, cold HI, MCs, magnetic fields and collisionless components of dark matter and stars. 
Further, in addition to turbulent and thermal gas pressure, magnetic field, cosmic rays and 
 radiation can play an important role under certain conditions. 
We simplify the analysis by considering a two component disk of 
 stars and gas. Treating the stellar component as a fluid coupled  to the gas exclusively via  gravity,  \cite{Jog84} present and analyze  the relevant equations governing the stability criterion in this case. 
 Based on this analysis,   \cite{Romeo2011} \citep[see also][]{Romeo2013} proposed the 
 following convenient approximation  for the Toomre parameter of such a   two fluid disk of gas and stars, 
\begin{equation}
\label{eq:Qeff}
\frac{1}{Q}=\frac{W}{Q_> }+\frac{1}{Q_< }
\end{equation}
where  $W=2\sigma_*\sigma_g/(\sigma_*^2+\sigma_g^2)$, $Q_>=\max(Q_*,Q_g)$ and $Q_<=\min(Q_*,Q_g)$.  Quantities with the asterisk refer to stars. The expression for $Q$ implies that $Q\le Q_<$, i.e. the disk is always unstable if the Toomre parameter of one of its components is less than unity.
We propose an effective velocity dispersion/sound speed for the composite system of gas and stars as
\begin{equation}
\label{eq:ceff}
\frac{1}{c_s}=\left(\frac{1}{Q_g \sigma_g }+\frac{1}{Q_* \sigma_*}\right)\frac{Q_g Q_*}{Q_g+Q_*}\; .
\end{equation}
Note that $c_s$ does not depend on $\kappa$. In \S\ref{sec:tests} 
we show that this is a reasonable ansatz.

An eigenmode  of the coupled perturbations in the two fluids  varies  with time as $\exp(i\omega t)$ where $\omega=\omega(k)$ depends on the wave number, $k$, of the eigenmode. 
Gravitational growth of modes with sufficiently large $k$ is suppressed by pressure, while angular momentum due to the rotation of the disk inhibits the growth of modes with sufficiently small $k$. Thus, collapse is possible for modes in a limited range of wavenumbers.
The eigenmode with the smallest  $\omega^2$, is termed here the least stable mode. 
Let $k_1$ and $\omega_1=\omega(k_1)$, be the wavenumber and frequency of this mode. 
Linear perturbation theory yields 
\begin{equation}
\label{eq:tgrow}
\omega_1^2=\kappa^2(1-Q^{-2}) \; .
\end{equation}
Given $Q$, this expression is exact in linear theory.
Corresponding to the  ansatz above  for $c_s$,   we  numerically demonstrate in  \S\ref{sec:tests} that
\begin{equation}
\label{eq:kone}
k_1\approx \frac{\kappa}{Q c_s}\; .
\end{equation}

Since $Q\sim \kappa$, the wavenumber  $k_1$ is actually independent of $\kappa$. 
Further, combining the two relations above, we find 
\begin{equation}
\omega_1^2\approx (c_s k_1)^2 \left(Q^2-1\right) \; ,
\end{equation} 
implying that  $\omega_1$ depends 
on $\kappa$ only via $Q$ in the term $(Q^2-1)$. 
For $Q<1$, $\omega_1^2<0$ and  the least stable eigenmode grows  exponentially with  a timescale $t_1=1/\sqrt{-\omega_1^2}$. 
In \S\ref{sec:tests} we demonstrate the validity of  the approximate relations in Eqs.~\ref{eq:ceff}--\ref{eq:kone} in relation to the full expressions in \cite{Jog84}.

It has been suggested  that  the stability criterion of a mode with a wavenumber $k$ in a turbulent medium should depend on the velocity dispersion generated by eddies with  sizes $\lsim \; 1/k$  \citep{Bonazzola1987,Romeo2010}. We will focus here on modes with $k$ in the vicinity of $k_1$ and assume that turbulence by gravitational collapse and SN feedback is generated on scales $\lsim  1/k_1$.

\section{Vertical balance}
The hydrostatic equation for vertical balance  between gravity and pressure of the gaseous component of the disk
is approximated as 
\begin{equation}
\label{eq:hydrofull}
\frac{1}{\rho^z_g}\frac{\dd p^z}{\dd z}=g_{g}+g_{*}+g_{h}\; .
\end{equation}
where  $\rho^z_g$, and $p^z$ are gas density and pressure, and 
 $g_{g}$, $g_{*}$ and $g_{h}$ denote  the vertical gravitational  force field, respectively,  
due  the gas, stellar component and the DM halo. In this equation all quantities are a function of the height, $z$, from the 
midplane of the disk. A detailed analysis of vertical disk balance can be found in \cite{McKee2015}.

For a  spherical DM halo with a circular velocity $v_{h}(R)$, we make the approximation  
$g_{h}=\Omega^2 z$, where $\Omega=v_{h}/R$ is the angular velocity  \footnote{Here $\Omega$ refers to the DM halo only.}.
The solution to  Eq.~\ref{eq:hydrofull} requires $g_*$ which in turn must be derived from the vertical distribution of the stellar component. Assuming that both the gaseous and stellar components are isothermal with different velocity dispersions, {numerical}
solutions to  Eq.~\ref{eq:hydrofull} can easily be found. 
Based on these solutions, we extract a convenient approximation   relating the 
midplane pressure, to the stellar and gas surface densities and velocity dispersions.
Let $p=p^z(z=0)$ and  $\rhog=\rhog^z(z=0)$  be
 the gas pressure  and density  in  midplane of the disk.
 The approximation is 
 \begin{equation}
 \label{eq:hydroproj}
p=\frac{\pi}{2} G \Sigma_{g} \widetilde \Sigma+
\frac{\Omega^2}{2 \pi}\frac{\Sg^2}{\rho_g}\; , 
\end{equation}
where  
\begin{equation}
\label{eq:wide}
\St =
\begin{cases}
\Sg+\frac{\sg}{\sast}\Sast \quad \textrm{for} \quad \sast \geq \sg\\
\Sigma_{tot}+\left(1-\frac{\sast}{\sg}\right)\Sast\quad \textrm{for} \quad\sast<\sg\; .
\end{cases}
\end{equation}
This is a generalization of the expression found by \cite{Elmegreen1989} neglecting  halo gravity and only 
for the case  $\sg<\sast$. 
The approximation in Eq.~ \ref{eq:hydroproj}  agrees with the full numerical solution within  10\% accuracy. 

Defining the disk height as $H=\Sg/\rhog$\footnote{For the forms $\rho(z)=\rhog \exp(-z/z_h)$ and $\rho(z)=\rhog \exp(-z^2/2z_h^2)$, we have $H=2z_h$ and $H=\sqrt{2\pi} z_h$, respectively.}, it follows from  Eq.~\ref{eq:hydroproj} that
 \begin{equation}
 \label{eq:dHwh}
H=\frac{2H_0}{1+\sqrt{1+\frac{2}{\pi} \left( \frac{\Omega H_0}{ \sg }\right)^2}}  \; ,
 \end{equation}  
where 
\begin{equation}
 \label{eq:dHnoh}
H_0=\frac{2\sigma_g^2}{\pi G \St}\; 
\end{equation}
is the disk height if the halo gravity is neglected, i.e. for self-gravitating disks as defined in \cite{Benitez-Llambay2018}.

According to  Eq.~\ref{eq:hydroproj}, the halo gravity is negligible if $
\pi^2 G  \widetilde \Sigma>\Omega^2 H,
$
which, upon taking $H\approx H_0$,  is equivalent to 
\begin{equation}
\widetilde Q_g\equiv \frac{\sqrt{2} \Omega \sigma_g}{\pi G\St}<{\sqrt{\pi}}.
\end{equation}
Thus, in agreement with \cite{Benitez-Llambay2018}, this inequality  implies that self-gravitating gas only disks  ($\St=\Sg$)  tend to be Toomre unstable.

For a self-gravitating 
gas-only  disk, we also 
find that the disk height is proportional to the inverse of the wavenumber of the least stable mode \citep[c.f.][]{Goldreich1965,Elmegreen2002}, 
\begin{equation}
\label{eq:Hkratio}
H k_1= 2\left(\frac{\sg^z}{\sg^R}\right)^2\; ,
\end{equation}
where we have taken $H=H_0$ from Eq.~\ref{eq:dHnoh} and $k_1$ from Eq.~\ref{eq:kone}
for the  gas-only case. In this expression, we distinguish between the velocity dispersion in the radial and vertical directions denoted, respectively, as $\sg^R$ and 
$\sg^z$. This is done for illustrative purposes only as we impose isotropy  $\sg^R=\sg^z=\sg$ throughout the paper.\footnote{For isotropic velocity dispersion,  vertical balance allows us to express the single fluid Toomre parameter 
 in terms of $\kappa$ and the midplane density, $\sqrt{2/\pi}Q=\kappa/\sqrt{G\rhog}$. }
In \S\ref{sec:tests}, we show that $H\sim 1/k_1$  for a composite disk with $Q<1$ 
and confirm that halo gravity has a substantial effect on the disk height only for $Q>1$.
Therefore, for $Q<1$, all relevant length scales are comparable. This is  important for  understanding  energy dissipation by means of turbulence cascade.

\section{Model star formation recipes}
We write the SFR per unit area as 
\begin{equation}
\label{eq:sfrtsf}
\Sfr=\epsf \frac{\Sg}{\tsf} \; ,
\end{equation}
where $\tsf$ is the timescale for gas to collapse  into stars and  $\epsf$ is the star formation efficiency parameter representing the fraction of $\Sg$  that actually turns into stars over this time scale. The efficiency parameter, $\epsf$ is an input parameter in our model.

We  consider two recipes for determining the time-scale, $\tsf$. 
The first, termed here the $Q$-recipe, is proposed by  \cite{Wang1994} and is based on Toomre instability. 
We take  $\tsf^{-1}=0$ for $Q>1$ and 
\begin{equation}
\label{eq:tsfQ}
\tsf^{-1}=|\omega_1| \; ,
\end{equation}
for $Q<1$. 

In the second recipe, the $\rho$-recipe, $\tsf$ is independent of $Q$ and is fixed by the local midplane density, 
\begin{equation}
\label{eq:tsfrho}
\tsf^{-1}=\sqrt{G\rhog} \; .
\end{equation}
This is the recipe proposed by \citep{Leroy2008}, however, in their case the effect of the halo gravity was not included. 
For a self-gravitating disk ($H=H_0$) this recipe yields 
$\Sfr\sim \Sg^2 (\St/\Sg)\sg^{-1}$ \citep[c.f.][]{Leroy2008}.
Also, for a self-gravitating gas-only  disk in vertical hydrostatic balance, $\sqrt{2 \pi} Q=\kappa/\sqrt{G\rhog}$, giving 
  $\sqrt{G\rhog}/\omega_1=\sqrt{2\pi}|1-Q^2|^{1/2}$ for the ratio  between  timescales in the  $Q$- and $\rho$-recipes. 
In both recipes,  $\sg$ and $\sast$ are needed to derive $\tsf$. This is done by  
equating energy gain and loss terms in Eq.~\ref{eq:Eschematic}. 
In the next section, we specify the forms of the gain and loss terms.

\section{Gain and Loss terms}

Dense cool shells of gas created by SN remnants eventually merge into the ISM, delivering momentum and energy. 
A remarkably simple result from analytic calculations  and simulations is that the momentum injected into the ISM by a single SN is   $p_*\sim (1-5) \times 10^5\; \textrm{M}_\odot \; \kms$ with weak dependence on  ISM density
\citep[e.g.][]{Cioffi1988,Kim2015,Martizzi2015}. 
Following \cite{Krumholz2018},   the energy gain (per unit gas mass) by SN 
is 
\begin{equation}
\mathcal{G}_\textrm{SN}=\frac{\Sfr}{\Sg} \frac{p_*}{m_*} \sg 
\end{equation}
   where $m_*\sim 100\, \textrm{M}_\odot$ is the average gas mass required to produce a single SN and  so $\Sfr/m_*$ is  the number of SN per unit area per unit time. The expression assumes that the shell merges with the ISM when its speed is roughly $\sg$.
The expression can be inferred as follows. If the gas mass swept up by the remnant at the time of the merging is $M^\textrm{rem} $, then the kinetic energy delivered by a single SN remnant is 
$p_*^2/2M^\textrm{rem}$. Hence, the energy injection per unit time per unit mass  is 
$(\Sfr/m_*)p_*^2/2M^\textrm{rem}/\Sg$ which leads to the expression above if 
$p_*/M^\textrm{rem}=2\sg$.

{ As  gravitationally unstable perturbations  evolve,
their kinetic energy increases  at the expense of their potential. This offers a source of  heating  via gravitational instability \citep[][]{Bournaud2010}. The presence  of negative and positive generic density perturbations is associated  with a large scale  coherent pattern  of divergent and convergent material
motions.  
Subsequent non-linear growth inevitably leads  
to  orbit  crossing  of 
large-scale portions of disk material. 
 This boosts cloud-cloud collisions in a cloudy  disk, raising the velocity dispersion of the clouds. Cloud-cloud collisions are inelastic and 
a fraction of  the initial energy is  transformed into internal cloud energy (e.g. to be consumed for triggering inter-cloud turbulence). However, we assume here that a significant 
fraction of the energy available in coherent motions is also transformed into 
random motions of the clouds \footnote{Galactic shear and orbit crossing on large scales can generate  a solenoidal (vortical) velocity component.}. Thus, as suggested by \cite{Bournaud2010}, we assume that turbulent motions
driven by gravitational instabilities, contribute to the disk height and affect its  stability properties \citep{Elmegreen2003}. 
}

 A generic linear perturbation is formed by  a superposition of eigenmodes  with different wavenumbers, $k$.
We assume that most of the heating is due to the least stable mode, i.e.  for $k= k_1$. 
We write the gravitational gain term as 
\begin{equation}
\label{eq:gain}
\cGgr=\gamma |\omega_1 | \sigma_{_{\!Q=1}}^2\; ,
\end{equation}
where $\gamma$ is a constant and 
\begin{equation}
 \sigma_{_{\!Q=1}} = \pi G \kappa^{-1} \left(\Sg+\Sast\right)\; .
 \end{equation} 
The  stellar velocity dispersion $\sast$ is required  to derive $\omega_1$ and we take $\sast=\gamma \sigQ$. We  adopt the value  $\gamma=1$ ;  thus for   $\sg=\sigQ$ we obtain $Q=1$ according to Eq.~\ref{eq:Qeff}, and the gravitational heating switches off.

As mentioned earlier, we  assume  that 
most of the  net energy  deposition  from SN and gravitational instability is transformed into   turbulent motions in the gas disk\footnote{
Gravitational instability also heats  stellar disk. However, an  analysis of the eigenvectors of  growing mode perturbations reveals that for $\sast=\sigma_{_{\!Q=1}}>\sg$ 
the amplitude of the density fluctuations in the stellar disk is generally much smaller  than in the gaseous disk.}.

We express the energy  loss term  by means of a turbulence cascade as 
\begin{equation}
\label{eq:tloss}
\mathcal{L}_{t}=\etat \sigma_t^2\frac{\sigt}{L_{t}}\; . 
\end{equation}
where $\etat$ is a constant and   the turbulence dissipation timescale is 
proportional to $\sigt/L_\textrm{t}$ with $L_{t}$ being the scale of the largest eddy  at which turbulence energy is injected. 
For the $Q$-recipe, the natural choice is the scale of 
the least stable mode, implying  $L_t=1/k_1$. 
Further, in this recipe the star formation is expected to form in spatial associations coherent over a scale $1/k_1$. 
Therefore, this choice for $L_t$ is also appropriate for  SN-driven  turbulence.
For the $\rho$-recipe, the relevant  scale is the disk height and thus we adopt $L_\textrm{t}=H$. 
Assuming that turbulence tends to produce an isotropic velocity dispersion \citep{Bournaud2010},  we have seen  (Eq.~\ref{eq:Hkratio}, see also  Fig.~\ref{fig:Hrat}) that  $H^{-1}\sim k_1$ provided 
$Q<1$. Therefore, in principle 
 taking $L_t=H$ in the $Q$-recipe as well, would have produced roughly similar results as $L_t=1/k_1$. 
 
 We express the energy balance equation as 
\begin{equation}
\label{eq:GainLoss}
\frac{3}{2\Sg}\frac{\dd \Sg \sigt^2}{\dd t}=\epsf  \frac{\vsn \sg}{\tsf}+\gamma  |\omega_1 | \sigQ^2-\etat\frac{ \sigt^3}{L_
{t}} \; ,
\end{equation}
where $ \vsn=p_*/m_*=1000-5000\kms$ for $m_*=100\, \textrm{M}_\odot$. The velocity dispersion, $\sigt$, is determined by 
setting the l.h.s to zero, i.e. at the equilibrium state where gain and loss terms cancel each other. 
In the presence of a thermal floor,   $\sigth> 0$, the $Q$-recipe does not allow real valued  solutions for disk parameters (e.g. very low $\Sg$), yielding 
$Q>1$ for  $\sg=\sigth$ (i.e. $\sigt=0$).  Formally, this implies a  vanishing SFR. However, the disk cannot be maintained indefinitely at a finite $\sigth$ without  any star formation and eventually a lower $\sigth$ will be established, at least in the average sense.
 We will not be concerned with tuning $\sigth$, so that a solution for $\sigt $ is found.  
In the $\rho$-recipe  a real valued  solution can always be be found since   {a $Q$ threshold} is not a pre-condition for star formation.

An important issue to address is whether our solution is stable against small variations in $\sigt$.
Since in  both recipes, all quantities $\tsf$, $\omega_1$ and $L_t$ depend on $\sigt$, an analytic assessment of this question 
is straightforward  in  limiting cases of purely gaseous disks,  showing that the solution is indeed stable. 
For general cases, we have confirmed  that full numerical solutions of Eq.~\ref{eq:GainLoss} always yields 
stable solutions $\sigt(t)$ approaching  $\dd \sigt^2/\dd t=0$ at sufficiently  large $t$.

\cite{Efstathiou2000} describes energy loss by means of cloud-cloud inelastic  collisions and derives a dissipation rate (per unit mass)
of the form $\Sg^2\sg$ (omitting pre-factors needed for proper units), which, upon using vertical balance for a purely gaseous disk, becomes $\Sg\sg^3/(GH)$. This is quite different from the form adopted here. However, we note that the velocity dispersion used in the collision rate is likely different from the turbulent velocity dispersion used in Eq.~\ref{eq:tloss}.

\section{Results}

We apply the relations in the previous section for a set of random 
parameters generated as follows.
The halo parameter $\Omega$ is drawn from a uniform distribution from  $\Omega=0$
to $\Omega=0.067\kms \,\textrm{pc}^{-1} $. The maximum $\Omega$ here corresponds to $\vh=200\kms $  
at $R=3\kpc$. The gas surface density is generated from a lognormal distribution 
in the range $0.1<\Sg/\Sigunits<10^5$. {This covers  low surface brightness (LSB)  galaxies ($\Sg \lsim \,  \mathcal{O}(1) \Sigunits$), normal galaxies  ($\sim \, \mathcal{O}(1) - \mathcal{O}(100) \Sigunits$) to starbursts ($> \mathcal{O}(100) \Sigunits$)}.

The SN momentum parameter, $\vsn,$ is uniformly distributed  between $\vsn=1000/m_{100}\kms$ and $5000/m_{100}\kms$ where $m_{100}=m_*/(100\, \textrm{M}_\odot )$ and $m_{100}$ is also uniformly distributed between $0.7 $ and $1.7$. Finally, the surface density ratio $\Sast/\Sg$ is log-normally distributed between $0.2$ and $10$.

The choice of the distribution of these parameters is  arbitrary and is not intended to represent any realistic distribution. Its purpose is simply to give an indication of the 
sensitivity of the star formation law to parameter variation.

\subsection{The star formation law: $\Sfr$ vs $\Sg$}
\label{sec:results1}

\subsubsection{Observations }

We compare the results to the updated  star formation law derived by \cite{DeLosReyes2019} for normal non-starbursting spiral galaxies. 
Adopting the SK form $\Sfr=A\Sg^n$\footnote{ $\Sfr$ and $\Sg$ in $\textrm{M}_\odot\, \textrm{pc}^2\, \textrm{Myr}^{-1} $ and $\Sigunits$, respectively.}, using the linmix estimator \citep{Kelly2007} these authors have found  $\log A=-3.84 $ and $n= 1.41$  with $1\sigma $ errors of $\sim 0.09$ and $\sim 0.07$, respectively. 
The intrinsic   rms scatter of $\log \Sfr$ at a given $\Sg$ around the 
SK relation is $\sim 0.28\, \textrm{dex}$.
 In addition we will  refer to the starburst and  
and LSB samples of galaxies, respectively,   compiled by \cite{Kennicutt2021}
and \cite{Wyder2009}. The starburst and normal galaxies can reasonable be fitted with a single SK law with $n\sim 1.5 $ \citep{Kennicutt2021}. However, 
examined on its own, starbursts fall on an SK relation with 
$ \log A=-2.8 $ and $n=1.16 $, markedly shallower than the SK law for normal non-starbursting spirals \citep{Kennicutt2021}. The estimated intrinsic scatter in the vertical direction of this relation is $0.33\, \textrm{dex}$, compared to  $0.28$ for normal spirals. 
 \cite{Kennicutt2021} also offer a reasonable   global SK fit to  observations of normal spirals and starbursts, with $\log A=-3.95$ and $n=1.54$.
 
 While  the statistical errors
on $A$ and $n$ 
are  small,  various  observational analyses 
 lead to slightly different parameters, depending on the calibration of the SFR, assumptions on the initial mass function, extinction corrections etc \citep{DeLosReyes2019}. Here  
 we  simply adopt the values specified above. 
 
These observations of $\Sfr$ versus $\Sg $ are represented Fig.~\ref{fig:sfrQ}. The black asterisks, black dots and red plus signs correspond to starbursts, normal spiral and LSB galaxies, respectively. The shaded areas represent the derived SK laws
for normal (green) and starburst galaxies (pink) where the vertical thickness reflects the intrinsic scatter on each case. The blue solid line is the global SK fit ($n=1.54$) advocated by \cite{Kennicutt2021}. 

Normal spirals in the figure span the range  $\sim 1-170 \Sigunits$ in gas surface density with roughly similar contributions from neutral and molecular hydrogen. 
In contrast, gas in starbursts is predominantly molecular 
with 
$\Sg \sim 90-3\times 10^4 \Sigunits $. 
{As such, the gas surface density of  starbursts depends crucially on the conversion of CO intensities to molecular hydrogen mass \citep[c.f.][]{Kennicutt2021}. Here we use the starburst data plotted in  figure 4 in \cite{Kennicutt2021} corresponding to a Milky Way conversion factor. }

A few galaxies in the normal and starburst populations overlap at $\Sg\sim 100 \Sigunits $ with starbursts having $\Sfr$ almost an order of magnitude higher than normal spirals. 
As $\Sg$ is increased, more starbursts tend fall  inside the  green shaded area representing the SK law inferred from normal spirals.
In fact, a simple linear regression of $\log \Sfr$ on $\log \Sg$ for all  starbursts yields a slope of $0.98$,  while the slope for  starbursts with $\log \Sg>3.5$ is $1.39$, very close to $1.37$
derived for  normal galaxies using linear regression\footnote{These slopes do not take into account the intrinsic and observational  scatter and are  different from the linmax estimates. They are given  here just to illustrate 
the point that 
starbursts tend to lie on  SK for $\Sg$ above a certain cut.}. This reinforces the visual impression that above a certain cut in $\Sg$, starbursts seem to obey the SK law of  normal spirals. 
LSB galaxies are neutral hydrogen dominated with $\Sg\sim 2 -7 \Sigunits $ and $\Sfr $ typically lying below normal galaxies with similar $\Sg$, but with a significant overlap between the two populations. 

\subsubsection{Model prediction: vanishing thermal floor}
For each  random selection of the parameter set $\Omega$, $\Sg$ and  $\Sast$, we derive $\sg$ by solving  the energy balance  Eq.~\ref{eq:GainLoss} with $\dd \Sg \sigt^2/\dd t=0$. Then
$\Sfr$ is derived  from the prescription in Eq.~\ref{eq:sfrtsf} for the $Q$ and $\rho$-recipes.
For the $Q$-recipe we adopt $\etat=\etavQ$ and $\epsf=\epsvQ$ as the default values, while for the $\rho$-recipes, we take $\etat=\etavH$ and $\epsf=\epsvH$. For convenience, these values are listed in table \ref{tab:default}. We will explore  the sensitivity  to these parameters in \S\ref{sec:epseta}.

\begin{table}
\centering 
\caption{Default values}
\begin{tabular}{ |l|c|c| } 
\hline
& $Q-$recipe & $\rho-$recipe\\
\hline
$ \etat$ & $\etavQ$ & $\etavH$ \\
$ \epsf$ & $\epsvQ $ & $\epsvH $ \\
\end{tabular}
\label{tab:default}
\end{table}

\begin{figure}
\includegraphics[width=.5\textwidth]{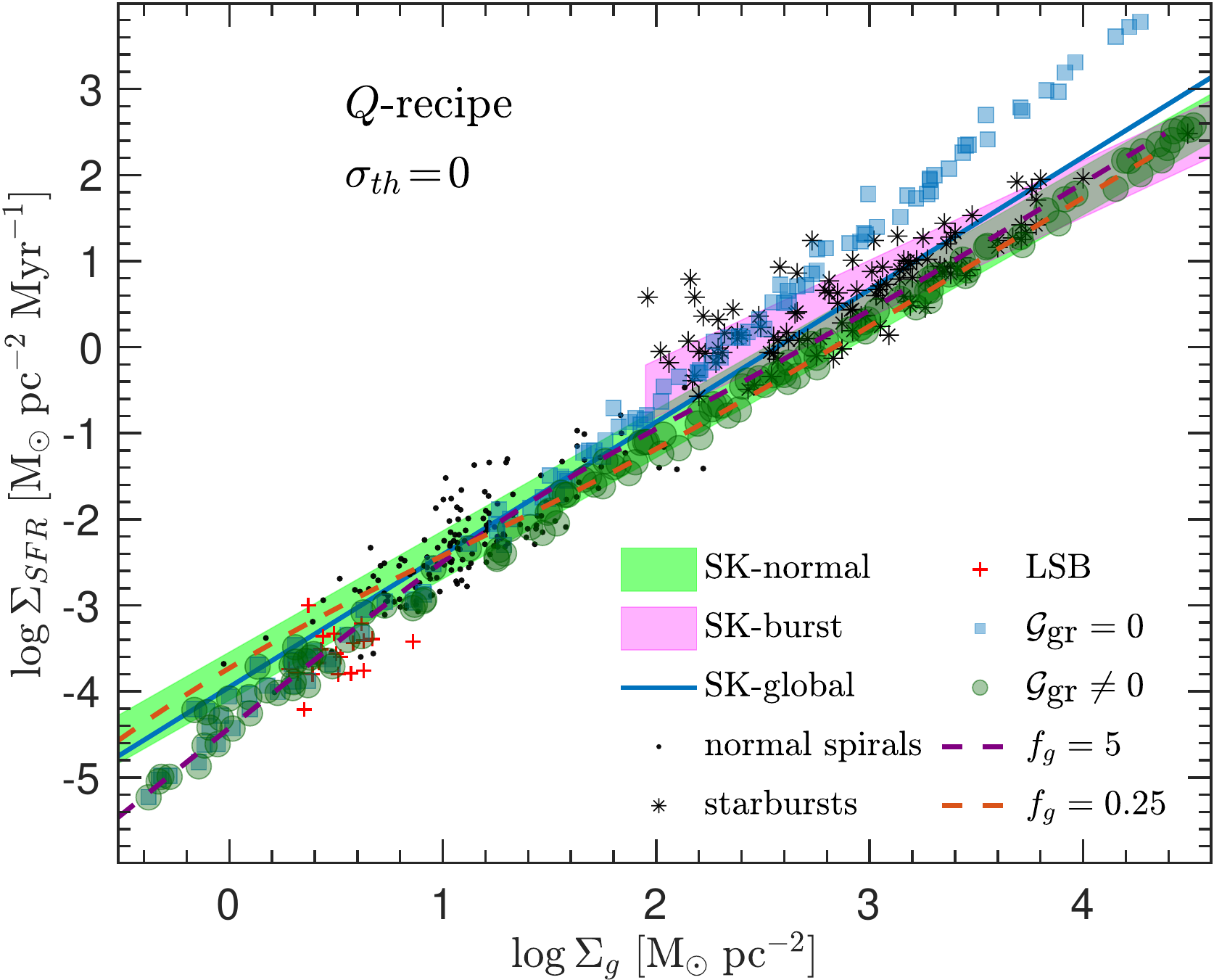} 
\vskip -0.8cm
\includegraphics[width=.5\textwidth]{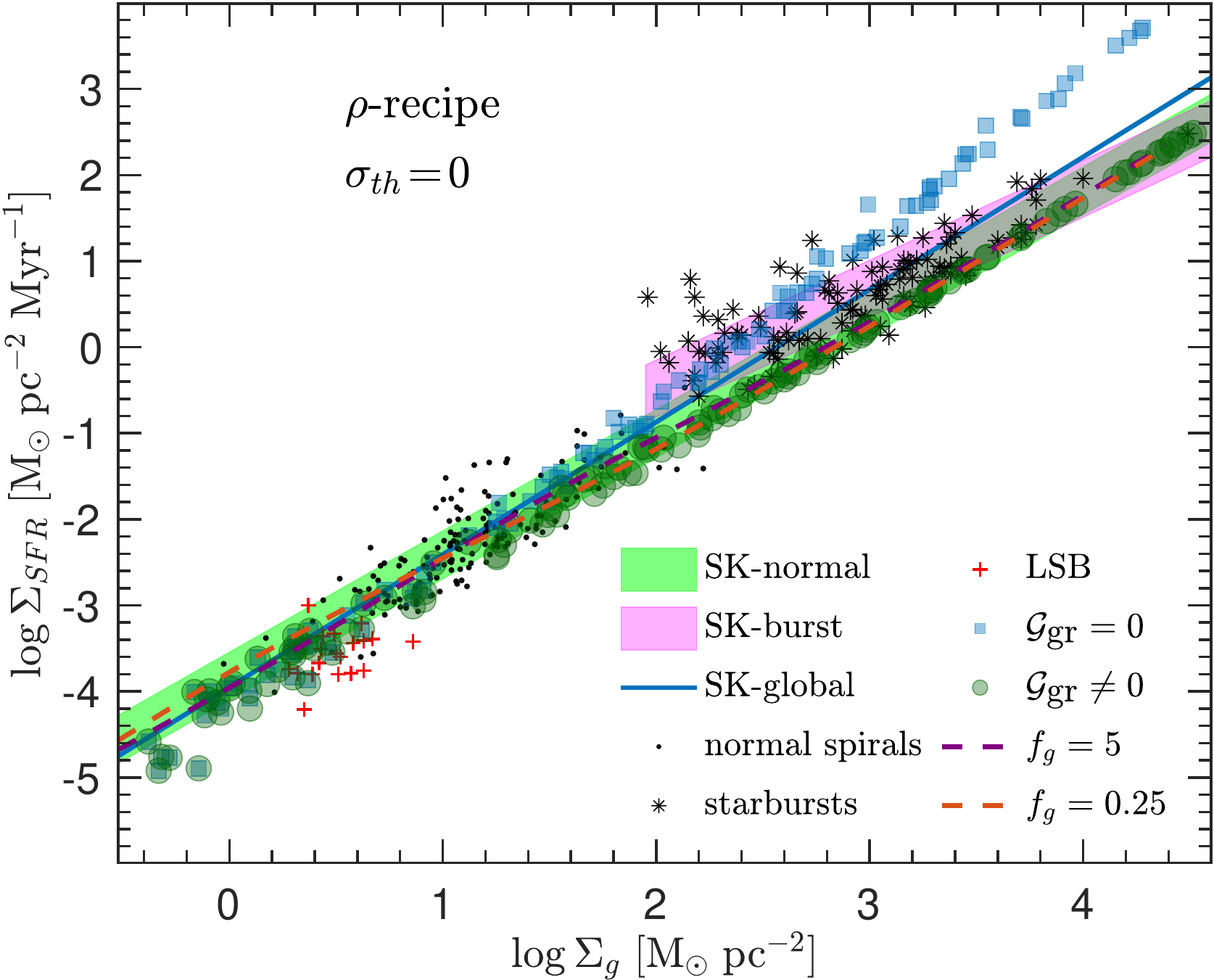} 
\caption{The star  formation rate per unit area versus the gas surface density for 100 sets of random parameters generated as described in the text. Blue squares are obtained with only SN feedback, while the green circles
also include gravitational heating. The dashed lines illustrate the dependence on the ratio $f_g=\Sg/\Sast$.
Green and pink shaded areas represent  the SK laws inferred, respectively, from observations of 
normal spirals  \citep{DeLosReyes2019} and starbursts  \citep{Kennicutt2021}. Their widths corresponds  the rms intrinsic scatter around the inferred law. 
Black dots and asterisks show  the actual measurements of 
normal spirals and starbursts. Red plus sign symbols  show  the LSB galaxy sample
of \citep{Wyder2009}.
The top panel and bottom panels correspond to  the $Q$- and $\rho$-recipe 
as indicated in the figure.
For comparison, the solid blue line is a global SK fit to the normal and starburst galaxies, with $n=1.54$.
The bottom panel is the same as the top but for the  $\rho$-recipe. There is a marked deviation of the blue points at  $\Sg\approx \Sgth= 50 \Sigunits$.}
\label{fig:sfrQ}
\end{figure}

In Fig.~\ref{fig:sfrQ} we examine our model predictions  in relation to these data sets.
The  top and bottom panels plot,  respectively, the $Q$-recipe and  $\rho$-recipe predictions for   $\Sfr$ versus $\Sg$ for 100 sets of random parameters  where we set the  floor velocity dispersion to zero, $\sigth=0$. 
The  blue squares are  obtained without the inclusion of  gravitational heating 
in Eq.~\ref{eq:GainLoss}  (i.e. with $\cGgr=0 $). 
 For both recipes, the blue squares  agree reasonably  well with the SK  law  for normal spirals  at 
$\Sg\lsim  \Sgth =50\Sigunits$  (i.e. $\log \Sg \lsim 1.7$), but    overpredict the SFR predicted from this law  at  higher surface densities.  A linear regression model of  $\log \Sfr$ on 
$\log \Sg$ for the blue points yields  slope of $1.9 $ and $1.85$, respectively, for the $Q$- and $\rho$-recipe. This is substantially steeper than the SK slope $n\sim 1.4$.
The blue points, however, are in reasonable agreement with most of the starbursts (asterisks) in the surface density range 
$\Sg \sim 100-300 \Sigunits$ (i.e. $ \log \Sg \sim 2-2.5   $), but they significantly deviate upward   from the data at higher surface densities.  At $\Sg \sim 10^4 \Sigunits $, they  overestimate the starburst $\Sfr$ by more than an order of magnitude.

 The green circles correspond  to solutions of the energy balance equation including $\cGgr$ with $\gamma=1$, in addition to SN feedback.
Gravitational heating 
has little effect 
at $\Sg\lsim \Sgth$, as indicated by the overlap of the green circles and blue squares in each  panel. At higher surface densities, 
gravitational heating becomes dominant (see also Fig.~\ref{fig:G}), bringing down the SFR required to maintain energy 
balance  to a better agreement with the observations. 
Note that in the $Q$-recipe, vertical pressure-disk weight is not used. 
This is in contrast to the $\rho$-recipe where the vertical balance equation is explicitly needed to compute $L_t=H$ and 
$\tsf\sim \rhog^{-1/2}$.

Gravitational heating creates a tendency  for flattening of $\Sfr$ versus $\Sg $ in the $Q$-recipe (green circles, top panel) at $\sim 100 \Sigunits$. The $\rho$-recipe (bottom) also exhibits a similar but less pronounced flattening. In both recipes  a steeper dependence on $\Sg$ with a slope  consistent with  SK for normal galaxies emerges at $\Sg\gsim 10^3 \Sigunits $.  
This is reminiscent of the behaviour of the models of \cite{Krumholz2018}.

Let us examine  the conditions for negligible gravitational heating, i.e. $\cGgr\ll \cGsn $, in the $Q$-recipe where $\tsf^{-1}=\omega_1$.  Consider   a purely gaseous disk, where  the solution to the energy balance equation is 
\begin{equation}
\label{eq:Qlimit}
Q_\textrm{g}^2=\frac{1}{1+\left(\frac{\etat \pi G \Sg}{\epsf \vsn\kappa}\right)^2}\; ,
\end{equation}
implying  $Q_\textrm{g}\approx 1$ at sufficiently small $\Sg$ and satisfying 
\begin{equation}
\label{eq:Qcond}
\pi G \Sg \ll \frac{\epsf}{\eta}\vsn\kappa\;  .
\end{equation}
Now,   comparing the terms $\cGgr$ and $\cGsn$, we find that $\cGgr$ is subdominant  if 
\begin{equation}
\label{eq:SNdom}
(\pi G \Sg/\kappa)^2\ll \epsf \vsn \sg     \; , 
\end{equation} 
holds. 
At   $Q_\textrm{g}\approx 1$, we have  $\sg\approx \pi G\Sg/\kappa $ and the last condition  becomes 
\begin{equation}
\pi G \Sg \ll  {\epsf}\kappa\vsn\; . 
\end{equation}
Since $\etat >1$ this condition is weaker than (\ref{eq:Qcond}). For $\kappa =0.034 \kms \; \textrm{pc}^{-1} $ and the default $\epsf$ and $\etat$, we find  that SN heating prevails for $\Sg \lsim \Sgth$ (with $Q\approx 1$) consistent with the top panel  in Fig.~\ref{fig:sfrQ}.

The scatter in the SFR law in the $Q$-recipe is mostly due to the 
variation in the ratio $\fg=\Sg/\Sast$,  as indicated by  the difference between the two dashed lines representing $f_g=0.25$ (amber ) and $f_\textrm{g}=5$ (purple) with fixed $ \Omega=0.034\kms \; \textrm{pc}^{-1} $ and $\vsn=3000\kms$.
The scatter in the $\rho$-recipe is tighter due to the weaker dependence on $f_g$ in the relevant expressions. 
The scatter of the observational data  includes observational errors and therefore cannot be directly compared to the scatter of the model predictions. 
Still, the rms of the intrinsic scatter of individual galaxies  at a given $\Sg$
as represented by the vertical extent of the shaded areas is larger than the 
scatter in the model prediction. The rms intrinsic scatter in $\log \Sfr$ for the $Q$- and $\rho$-recipes  with gravitational heating is $0.2 $ and 
$0.15$, to be compared  to $0.28$
of the data. But so far the model prediction did not include possible variations in the parameters  $\epsf$ and $\eta $ which we will  at a later stage.

\subsubsection{Model predictions: non-vanishing thermal floor}
Next we explore  effect of a non-vanishing $\sigth$.
Following \cite{Krumholz2018} we take  $\sigth$  to depend linearly the mass fraction of molecular gas  ranging from $0.2 \kms$ in the purely molecular phase to 
$5.4\kms $ for neutral hydrogen. From the normal spiral data in \cite{DeLosReyes2019}
we find that galaxies with $\Sg\lsim <3\Sigunits $ are dominated by neutral hydrogen 
while for $\Sg\gsim 30 \Sigunits $ the gas is primarily in molecular phase. 
In the middle range between these values  the fraction of gas in the molecular phase  
is nearly uniformly distributed between $0$ and $1$.
For a given $\Sg$ in our model we randomly pick a molecular gas fraction reflecting this observed distribution and then derive the corresponding $\sigth$. This yields  $\sigth=5.4\kms$  and $0.2\kms$, respectively, at the low and high ends of 
$\Sg$.

\begin{figure}
\includegraphics[width=.5\textwidth]{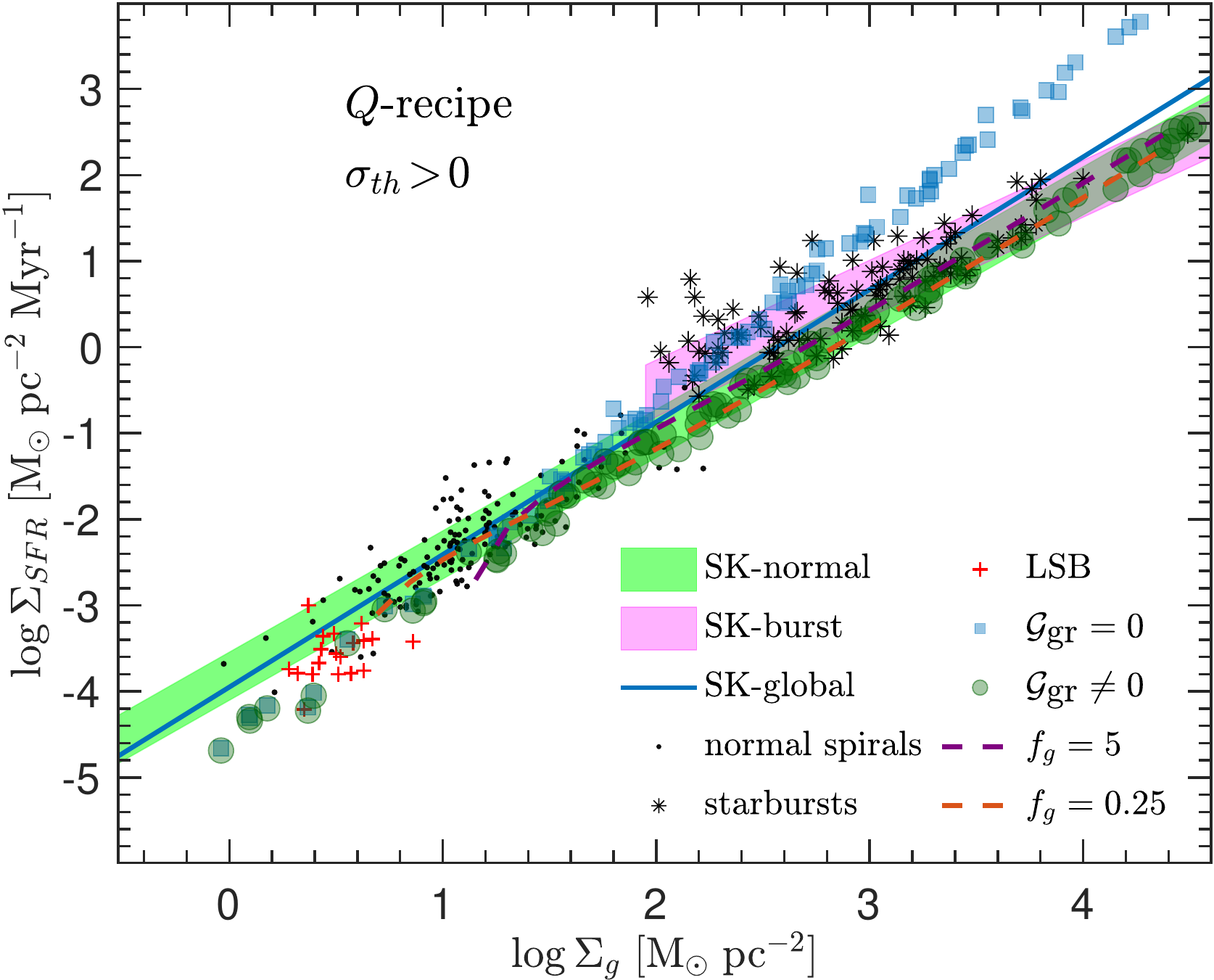} 
\vskip -0.8cm
\includegraphics[width=.5\textwidth]{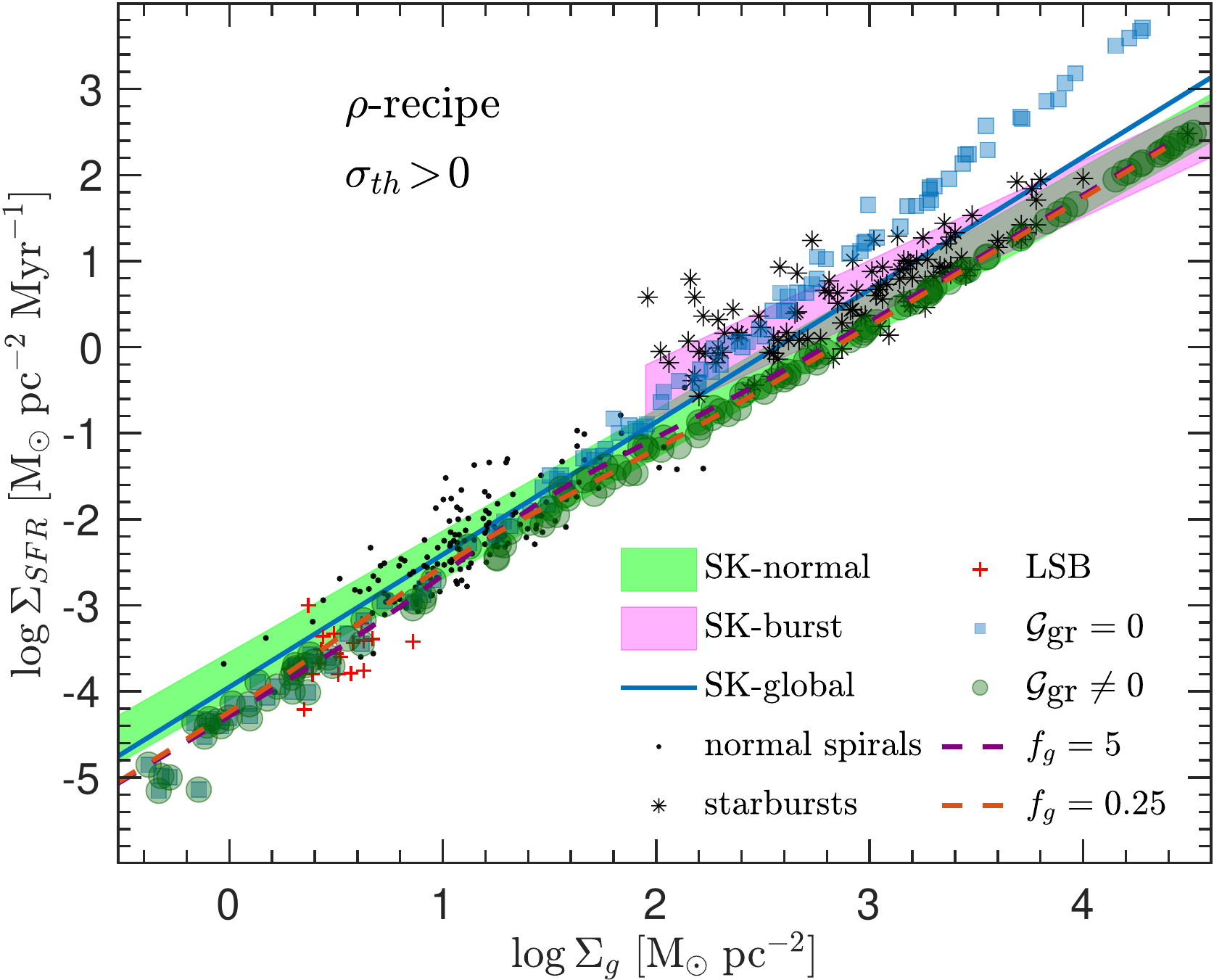} 
\caption{The same as the previous figure, but with a non-vanishing thermal floor  in the velocity dispersion.}
\label{fig:sfrQth}
\end{figure}

The results obtained with non-vanishing $\sigth$ are presented  in Fig.~\ref{fig:sfrQth}.
In both recipes,  the main difference with the previous figure ($\sigth=0$) is at low $\Sg$. 
In the top panel ($Q$-recipe) there are no  points at low $\Sg<1\Sigunits $ where the $\sigth$ floor pushes 
$Q$  above the threshold 
$Q=1$ for star formation. However, we also have run code with a larger number of sets of random parameters 
and found that some  points with this low $\Sg$ actually have non-vanishing but  low SFRs ($\log \Sfr\sim -5.5$). 
Those points are obtained for low  $\Sast$ and $\kappa$  
that just  happen  to yield $Q<1$. 
Spatially resolved observations of nearby galaxies probe this low  $\Sg$ range and they 
indeed exhibit suppressed SFRs relative to the SK of normal galaxies (green area in the figure) \citep{Bigiel2010}. 
Due to the weaker dependence of the scale height on $\sg$, the thermal floor has a less dramatic effect in the $\rho$-recip with a reduction by a factor of $\sim 5$
at $\Sg\sim 1 \Sigunits$. 

Appendix \ref{sec:depG} presents further analyses and numerical investigation in terms of  the depletion time scale, $\tsf/\epsf$, and the gain and loss terms.

\subsection{Velocity dispersion and Toomre parameter}
 
\begin{figure}
\includegraphics[width=.5\textwidth]{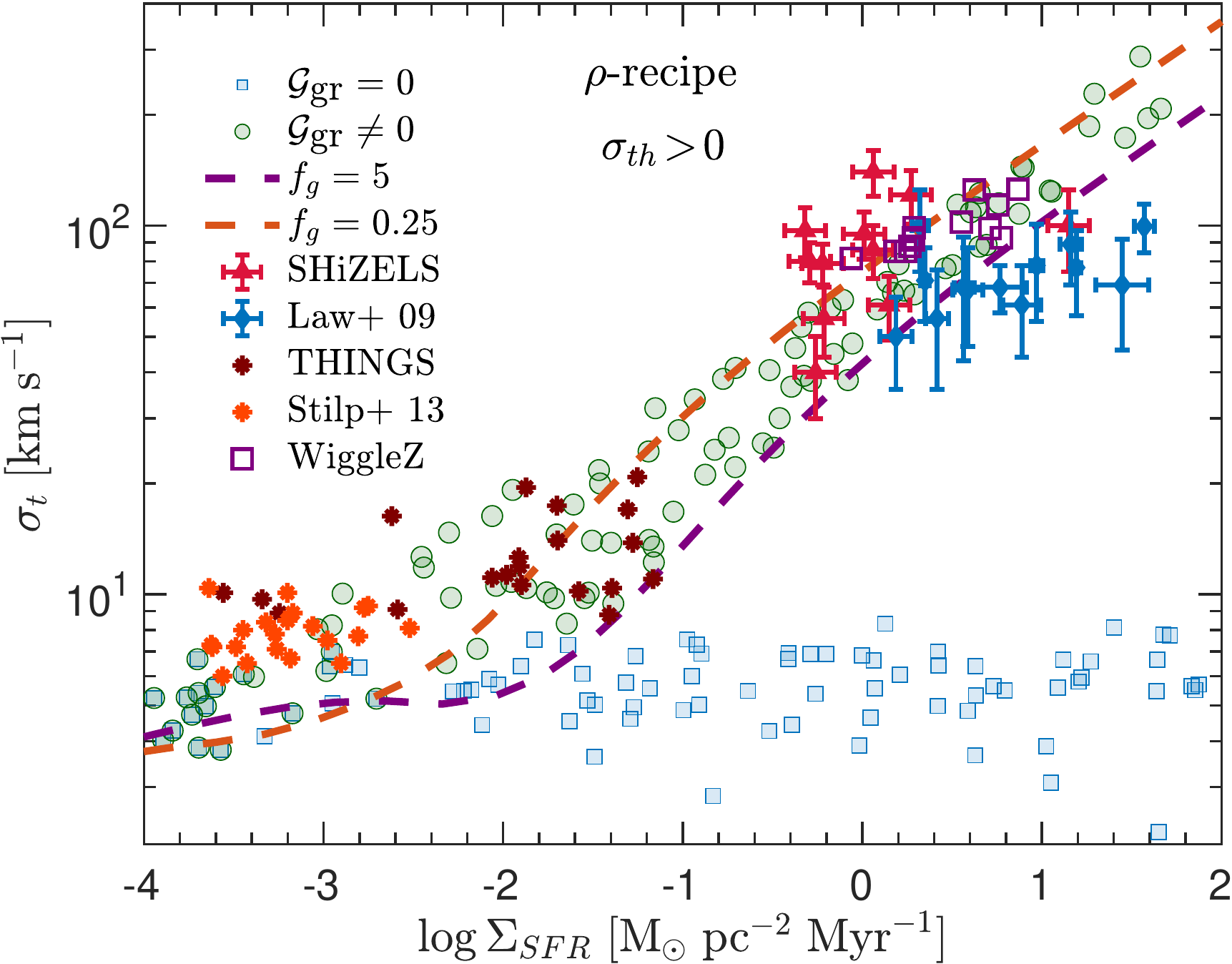}
\vskip 0.5cm
\caption{The turbulent component of the gas  velocity dispersion, $\sigt$,  versus $\Sfr$ for the $\rho$-recipe with a thermal velocity dispersion floor $\sigth=5\kms$. Circles and squares represent  solutions of the energy balance equation with and without gravitational heating, while the red triangles  correspond to data from  \citep{Swinbank2012}. The dashed lines correspond to gas to $\fg=\Sg/\Sast=0.25$ and  $5$, as indicated in the figure.}
 \label{fig:sigv}
\end{figure}

As emphasized by \cite{Krumholz2018}, models should be able to recover the observed $\sg$ in relation to the SFR. 
\cite{Krumholz2018} explore their model prediction for $\sg$ versus  the total star formation rate, $\dot M_*$ (total mass in stars formed per unit time).  Their detailed mass transport model allows them to integrate $\Sfr$ to obtain $\dot M_*$. In principle, we can augment our model with assumptions regarding the size of the disk in order to  estimate $\dot M_*$. However, we opt  to  restrict the analysis to the more robust 
model prediction for  the correlation of  $\sg$ versus $\Sfr$.
The results are presented in Fig.~\ref{fig:sigv} for the sets  of random parameters, as the green circles (including gravitational and SN heating) and blue squares (including only SN  heating) in the $Q$-recipe. Dashed curves correspond to two specific values for $f_g$, as in Fig.~\ref{fig:sfrQ}.  The figure shows  the turbulent component of the velocity dispersion, which should be added in quadrature to $\sigth$ to find $\sg$.  Similar results are obtained for the $\rho$-recipe and hence, for brevity, we show only the $Q$-recipe.

The difference between the predictions with and without  gravitational heating  is  pronounced at high $\Sfr$. 
As seen by the blue points in the figure,  SN heating alone leads to a nearly constant $\sigt$  over the whole plotted range of six orders of magnitude in $\Sfr$. 
Gravitational heating  boosts  $\sigt$ at high $\Sfr$, as indicated by the rise of the green circles for  $\Sfr\gsim 10^{-2} \Sigunits\, \textrm{Myr}^{-1} $ (corresponding to $\Sg\gsim \Sgth$).

\cite{Swinbank2012} present measurements of $\sg$ versus $\Sfr$  in a   sample of nine star-forming galaxies at redshifts $z=0.8-2.2$. 
They derive the velocity dispersion from star forming clumps in the SHiZEL galaxies.  These authors  also find that 
these galaxies 
  follow an SK law with parameters consistent with the local relation. Therefore, a  comparison of  their measurements with the results obtained here is appropriate.  The data points are shown as the red crosses in the figure with the bars representing 
 the errors provided  in \cite{Swinbank2012}. Although, the data points lie somewhat above the theoretical predictions for 
 the gravitational heating case, the overall agreement is encouraging. 
The velocity dispersion obtained  with SN heating alone, is markedly  below the  observations.

{  Also plotted in the figure are  data used in \cite{Krumholz2018} whenever the sizes of the  star-forming regions are available. Once this information is available we estimate $\Sfr$.
For the galaxies in $z\sim 2-3 $ sample of \citep{Law2009}, the area of nebular emission is provided in their table  3. The corresponding data is shown as blue diamonds, where the horizontal error bars represent the  uncertainties in the area only and not in the observed $\dot M_\ast$. The vertical error bars on $\sg$ are taken from their table 5.
For The HI Nearby Galaxy Survey 
\citep[THINGS, ][]{Ianjamasimanana2012}, we take $\dot M_\ast$ and $l_{SFR} $ from table 4 in \cite{Leroy2008} and estimate  $\Sfr \approx 0.26\dot M_\ast /\pi l_{SFR}^2$ where the factor $0.26 $ accounts for the fraction of  SFR within the scale radius $l_{SFR}$ of their  exponential fits to the radial SFR profiles.  We have found by visual inspection that these  estimates of the $\Sfr$ are a reasonable fit to the averages of the $\Sfr $ profiles plotted  in the appendix of \cite{Leroy2008}. 
For the sample of local dwarf galaxies of \cite{Stilp2013a}, 
the values of $\Sfr $ are provided in their table 3.
The relevant data for the  WiggleZ ($z\sim 1.3 $)  galaxies 
are available in \cite{Wisnioski2011}.  We follow \cite{Krumholz2018} and estimate $\dot M_\ast$ from the H$\alpha$ luminosity using the relation in \cite{Kennicutt2012}. We derive $\Sfr$ based on the half-light radius of H$\alpha$ emission given in table 4 of \cite{Wisnioski2011}.
}

{Although we do not have  error bars available for  all data points, 
there can be up to a factor of two ambiguity in the derivation of $\Sfr$.  There is a good general agreement  between the observations and the gravitational heating model predictions at  $\Sfr \gsim 10^{-3}\,  \Sigunits \, {\rm Myr}^{-1}$.
 At lower values of $\Sfr$, the models with and without gravitational  heating tend to underpredict the observed velocity dispersion, mainly the \citep{Stilp2013a} sample of local dwarfs and low mass spirals. 
Note that only the turbulent component is plotted for the model prediction. The addition of a thermal component of  $\sigth\sim 5\kms$ 
mitigates this discrepancy for some galaxies, but it is unlikely  to explain
the velocity dispersion of $\sim 10 \kms$ \citep{Stilp2013}.
Therefore, as discussed in detail in \cite{Stilp2013}, it is 
unclear how to account for high velocity dispersion seen in some galaxies in their sample of local dwarfs and low mass spirals.}

\begin{figure}
\includegraphics[width=.5\textwidth]{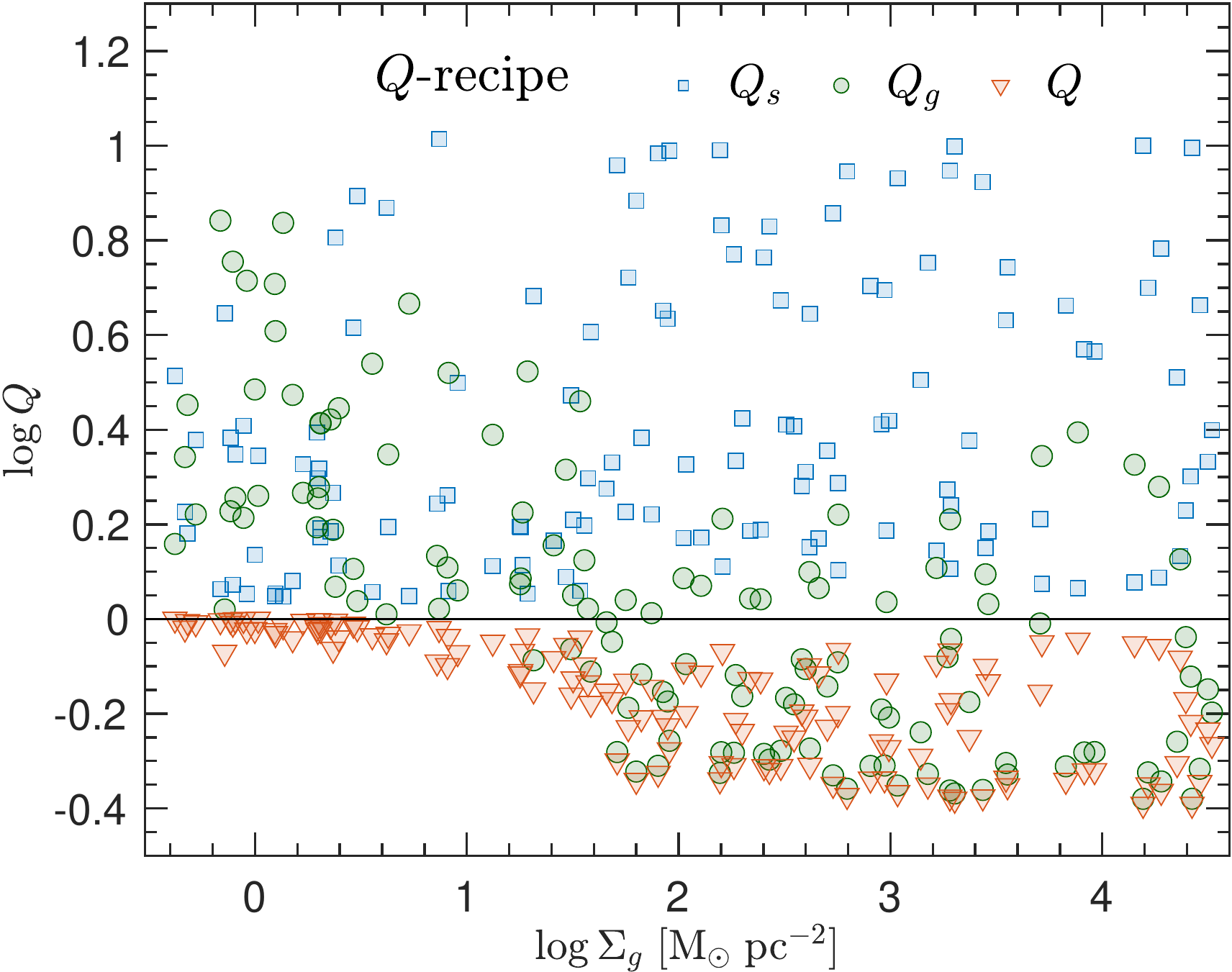}
\vskip -0.7cm
\includegraphics[width=.5\textwidth]{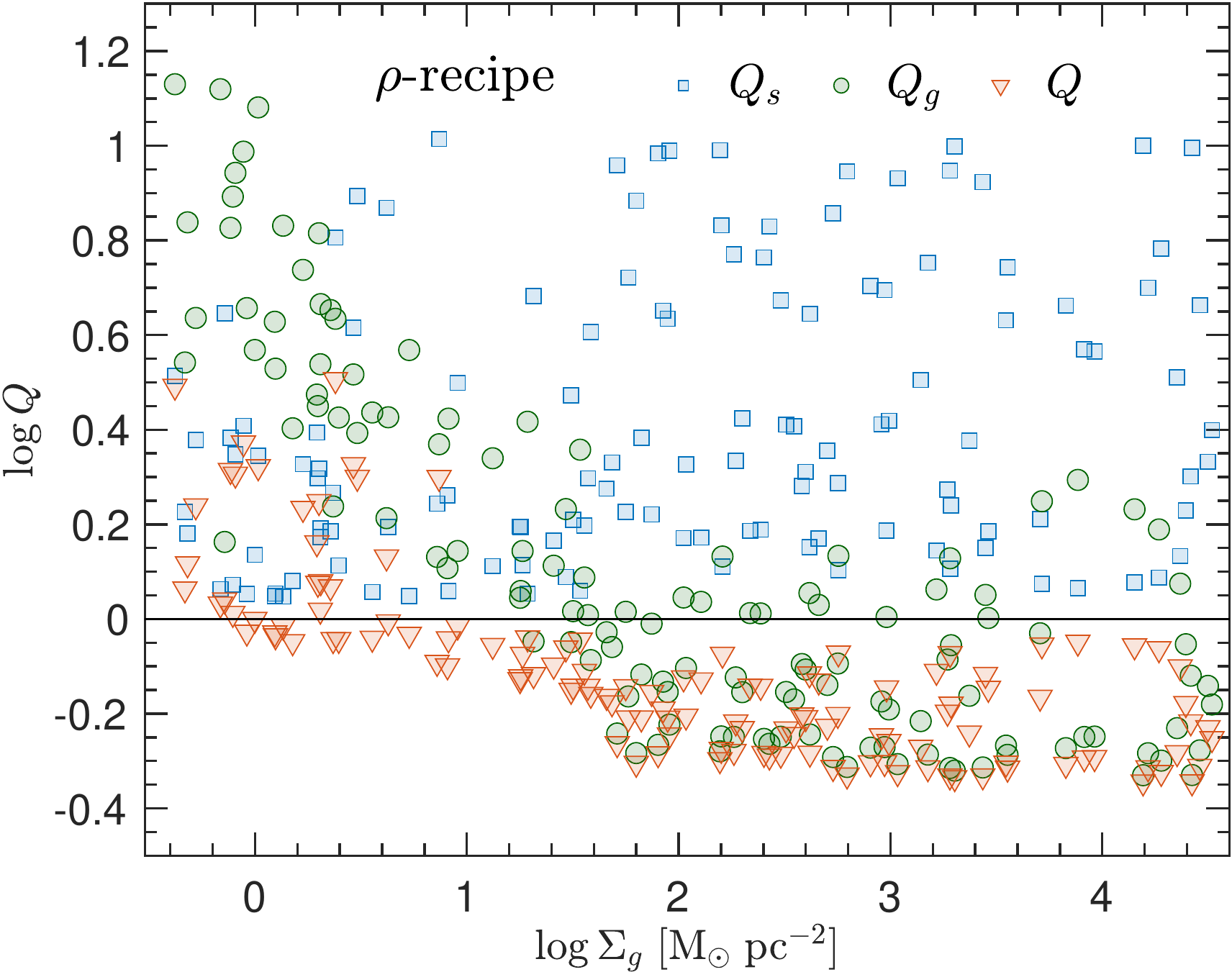}
\caption{The global, stellar and gas Toomre parameters obtained in the  two star formation recipes. Top and bottom panels are for the $Q$- and $\rho$-recipes, as indicated in the figure.}
 \label{fig:Q}
\end{figure}

Related to the velocity dispersion is  the  Toomre parameter. 
This is plotted in Fig.~\ref{fig:Q} for $\sigth=0$ for the two star formation recipes.
The results for $\sigth=5\kms$ are similar except that at low  $\Sg$ there are fewer  in the $Q$-recipe, while slightly larger values are acquired by $Q$  in the $\rho$-recipe. 

By construction, the stellar Toomre parameter, $Q_s$,  remains above unity in
both recipes. 
For the $Q$-recipe (top panel), the global $Q$ is always below unity since a stable disk ($Q>1$) neither forms stars nor allows for collapse of perturbations. Hence the heating terms vanish and no energy balance is possible for $Q>1$. There is a transition  in the distribution of $Q$ as a function of $\Sg$. Below  $\Sg \approx \Sgth$, both $Q_g$ and $Q_s$ are generally above unity, 
while the $Q$ values are concentrated around unity with little scatter at  $\Sg\rightarrow 0$. This is consistent with the expression in Eq.~\ref{eq:Qlimit} for $Q$ derived  for low $\Sg$.

Above $\Sgth$, both $Q$ and $Q_g$ spread   below unity reaching values as low as $0.3$.  
A detailed inspection of the results, reveals that points with low  $Q$ are associated with high $\Sg/\Sast$ ratios,    while  $Q\approx 1$ are obtained for disks dominated by the stellar component. 

The $\rho$-recipe (bottom panel) also shows a similar transition of $Q$ versus $\Sg$ reflecting the importance of 
gravitational heating at $\Sg\gsim \Sgth$. At  high $\Sg$, the results are very similar to the $Q$-recipe since the energy input from star formation becomes subdominant and energy balance is possible only for unstable disks. 
At low $\Sg$, star formation is possible for $Q>1$ in the $\rho$-recipe, as seen in the figure. 

{There are several observational estimates of the Toomre parameter in the literature. We emphasize that the value  $Q=1$ marking the stability transition is obtained under limited conditions   and should not  strictly hold for realistic cases due to deviations from these assumptions and 
large observational uncertainties.
\cite{Kennicutt1989} and \cite{Martin2001} measure the Toomre parameter using the expression for a single fluid without including the stellar component and assuming the same  velocity dispersion of  $6 \kms$ for all galaxies in their analysis. These measurements were consistent with the SFR related to a critical surface density defined by the Toomre parameter, but with large variations. 
\cite{Leroy2008}  provide estimates of the Toomre parameters for THINGS galaxies assuming a fixed gas velocity dispersion of $11\kms$ and  stellar velocity dispersion inferred  from vertical hydrostatic equilibrium. 
These authors obtain $Q$, $Q_s$ and $Q_g$ values that are markedly  above the stability regime advocated by \cite{Kennicutt1989} and \cite{Martin2001} with a large scatter.
They show that this is a reflection of different assumptions from \cite{Kennicutt1989} and \cite{Martin2001}  regarding the velocity dispersion and the conversion of CO emission into mass of molecular hydrogen.
Therefore,  we conclude that detailed comparisons of model $Q$ with observations are indecisive. }

\subsection{Scatter and Dependence on  $\epsf$ and  $\etat$}
\label{sec:epseta}

The analysis  so far referred to a single choice for the parameters $\epsf$ and $\etat$. Many factors play a role in fixing these parameters, e.g. the details of the disk morphology,  properties of star forming molecular gas, external tidal fields, and the DM distribution. 
Therefore, both of these parameters are likely  to vary among galaxies and even across  regions in the same galaxy. These variations can enhance the scatter in the predicted $\Sfr$.

\begin{figure}
\includegraphics[width=.5\textwidth]{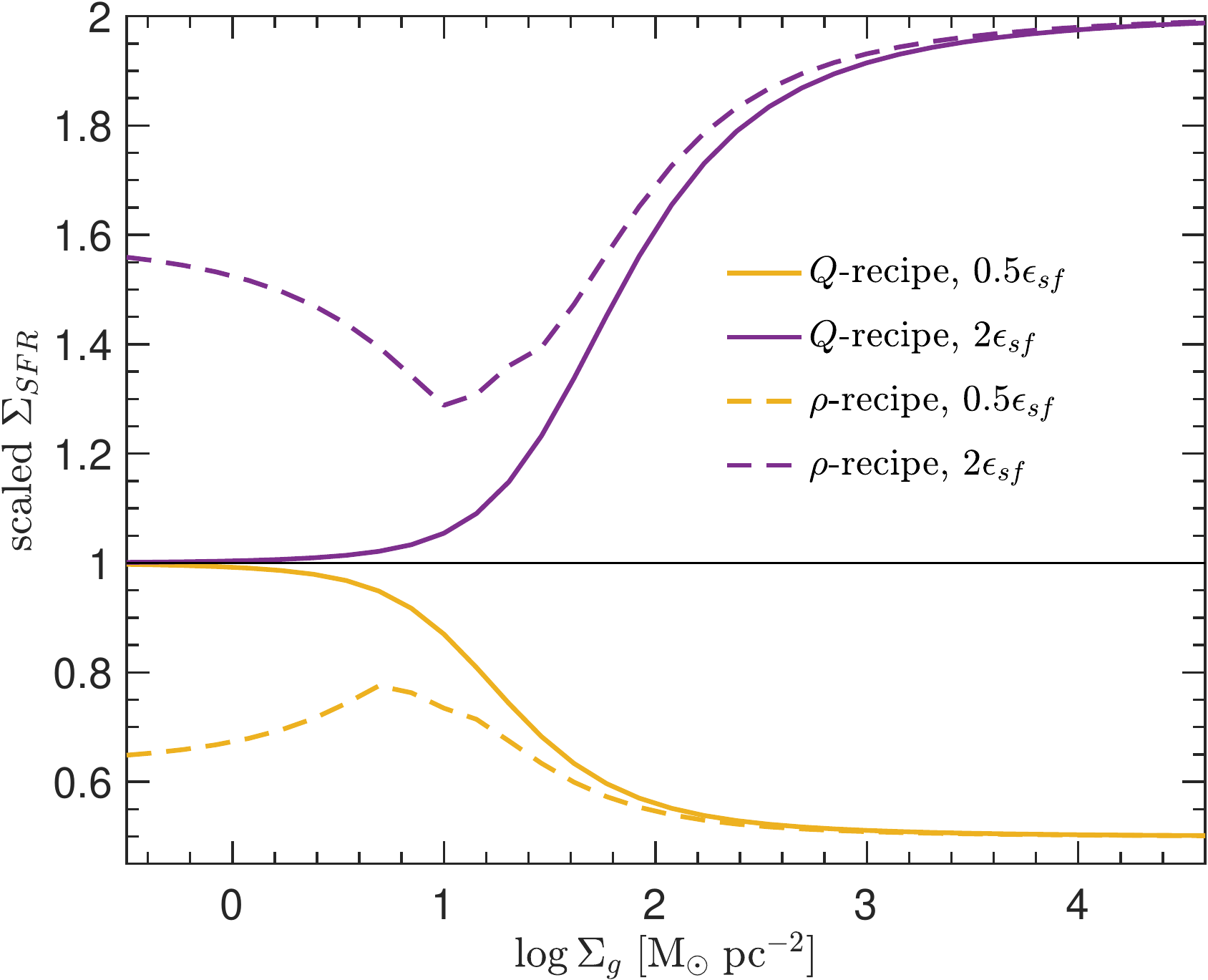} 
\vskip -0.8cm
\includegraphics[width=.5\textwidth]{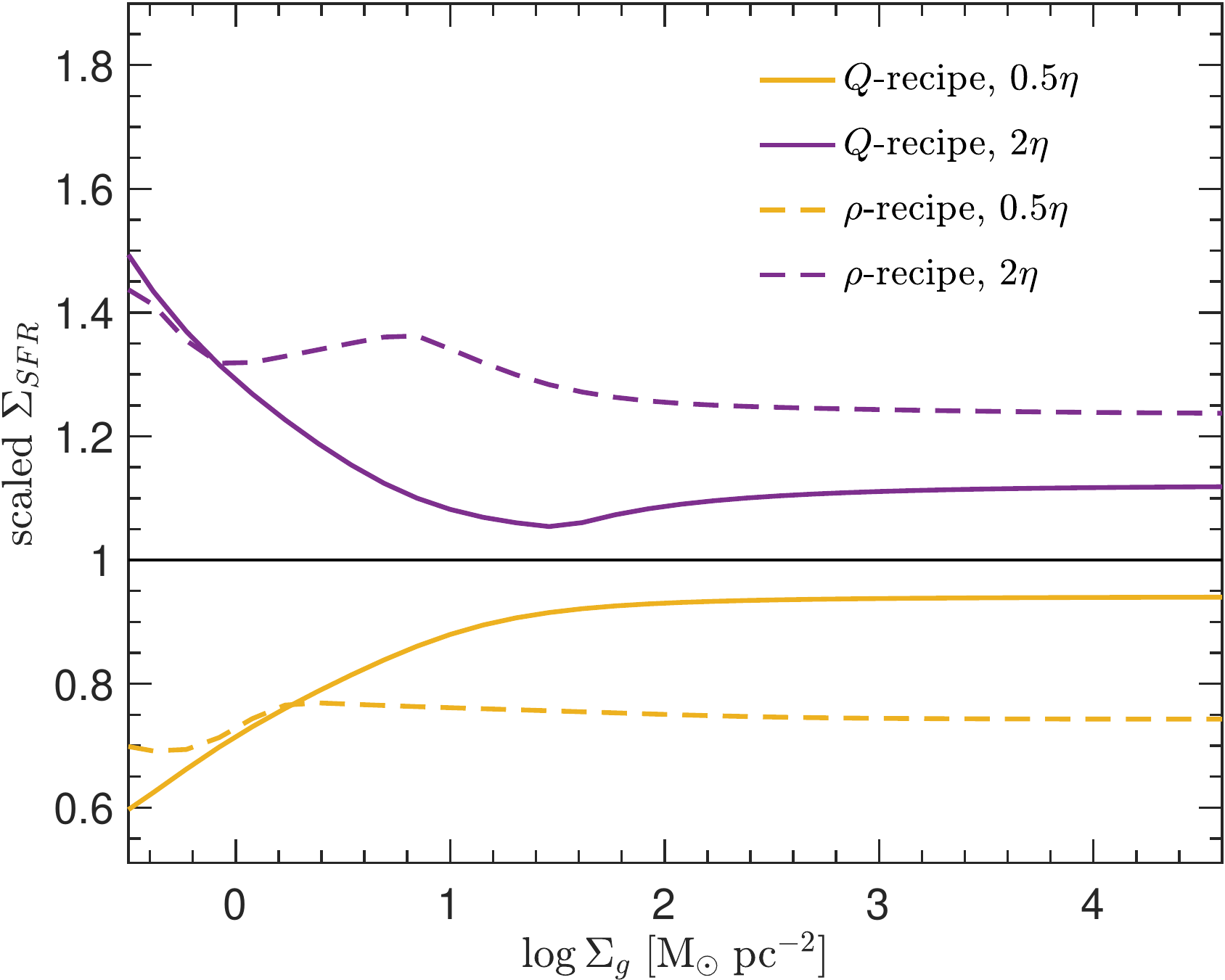} 
\caption{Dependence on the parameters $\epsf$ (top) and $\etat$ (bottom). Solid  and dashed lines  refer, respectively, to the $Q$- and $\rho$-recipe. Purple and yellow colors, respectively,  correspond to twice and one half the default values. }
 \label{fig:epsdep}
\end{figure}

{We  explore  the dependence of the $\Sfr-\Sg$ relation on $\epsf$ and $\etat$ in Fig.~\ref{fig:epsdep}. 
In both panels, we fix  $ \Omega=0.034\kms \; \textrm{pc}^{-1} $ and $\vsn=3000\kms$. 
The top panel refers to varying $\epsf$, while fixing $\etat$ at the default values
in each recipe (table \ref{tab:default}). 
The  $\Sfr$ curves in this panel  correspond to  $0.5\epsf$ (amber) and $2\epsf$ (purple) where here $\epsf$ is the default value in each recipe. 
Similarly, the bottom panel  probes  the sensitivity to $\etat$ with  $\epsf$ fixed at the default values.
In the two panels, we plot $\Sfr$ normalized by the   curve obtained from the default values of $\etat$ and $\epsf$.}

There is a strong dependence on $\epsf$ at $\Sg\gsim \Sgth$ in both recipes. At large $\Sg$, SN heating becomes  negligible and the velocity dispersion is mainly fixed by  a balance between  gravitational heating and  dissipation. 
Therefore, $\tsf$ is almost independent of $\epsf$ and $\Sfr=\epsf \Sg/\tsf$ varies linearly with $\epsf$  in the limit of very large $\Sg$ as 
seen in the top panel in the figure. 

In the regime of  $\Sg\lsim \Sgth$ where gravitational 
heating is subdominant,  the dependence on $\epsf $ is very weak for the $Q$-recipes. 
 This interesting result can be understood as follows. Eq.~\ref{eq:Qlimit}, appropriate for this recipe in the low $\Sg$ limit, implies 
 \begin{equation}
  \omega_1=(217 \textrm{Myr})^{-1}\left(\frac{\etat}{10}\right)\left( \frac{0.01}{\epsf} \right)\left(\frac{3000}{\vsn}\right)\Sg 
 \end{equation}
 where $\Sg$ in $\Sigunits$, $\vsn$ in $\kms$  and we have used $\omega_1=\kappa(Q^{-2} -1)^{1/2}$. The result is independent of $\kappa$.
 Therefore, 
\begin{eqnarray}
\label{eq:noeps}
\Sfr&=&\epsf \Sg \omega_1 \approx  \etat \frac{\pi G \Sg^2}{\vsn}\\
&=& 4.5\times 10^{-5}\left(\frac{\etat}{10}\right)\left(\frac{3000}{\vsn} \right)\Sg^2 \; ,
\end{eqnarray} 
which is independent of both $\epsf$ and $\kappa$. Although the quadratic dependence in this limit does not match the SK law on average, the scatter in the prediction due to varying $\Sg/\Sast$ is large at low $\Sg$ as seen in Fig.~\ref{fig:sfrQ}.

In contrast,  the figure shows that SFR in the   $\rho$-recipe in the limit of $\Sg\rightarrow 0$,  is sensitive to  
to $\epsf$   and implies that $\Sfr\sim \epsf^{1/2}$. This is consistent with the following analytic considerations. 
The dissipation scale in this recipe is the disk height $H$ and the SF time scale if fixed by $1/\sqrt{G\rho}$. The vertical balance equation Eq.~\ref{eq:hydroproj} can be used to relate these quantities to $\sigma$ 
in the energy balance equation Eq.~\ref{eq:GainLoss}. In the limit of very small $\Sg $, 
only the halo gravity is relevant on the r.h.s of Eq.~\ref{eq:hydroproj}, giving 
$\rho \sg=\Omega \Sg/\sqrt{2\pi}$. Using this in the energy balance equation without the inclusion of the Toomre heating term, we arrive at 
\begin{equation}
\sg=(2\pi)^{1/4}\frac{\epsf}{\etat}\frac{\vsn}{\Omega}\left(\Omega G \Sg\right)^{1/2}\; .
\end{equation}
After some algebra, we write the SFR in the low $\Sg$ limit in the $\rho$-recipe as 
\begin{equation}
\Sfr=(2\pi)^{-3/8}\left(\epsf \etat \right)^{1/2}\!\left(\frac{G\Omega^3}{\vsn}\right)^{1/2}\!\Sg^{3/2}\; .
\end{equation}
Thus we $\Sfr \sim \epsf^{1/2}$ as seen in the figure for small $\Sg$.

The bottom panel of Fig.~\ref{fig:epsdep}   explores the sensitivity  on the dissipation parameter $\etat$. 
In the limit of small $\Sg$, the dependence in both recipe is consistent with the expressions derived above, i.e. $\Sfr\sim \etat$ and 
$\etat^{1/2}$, respectively, for the $Q$- and $\rho$-recipes. 
At  large $\Sg$,  energy balance between the dominant gravitational  heating and dissipation determines $\sg$ and thus, indirectly, $\Sfr$. 
The dependence can in principle be derived analytically in both recipes. For example,  for an entirely gaseous disk in the $Q$-recipe,  $\Sfr \sim \etat^{1/2}$ at large surface densities which does not fit behavior of the solid curves in the lower panel in Fig.~\ref{fig:epsdep} due to the small gas fraction ($f_g=0.25$) adopted in this figure.

\section{Porosity}
 Global self-regulation by SN feedback alone can  be achieved if the fraction of volume, $f$, occupied by  hot gas in percolating SN remnants is appreciable but also not too large \citep{Silk1997,Silk2001,Li2015}.
The filing fraction is $f=1-\mathrm{e}^{-\cP}$, where the porosity, $\cP$, is the product of the  SN  rate (number per unit volume per unit time) times the 4-volume 
 of an  SN remnant at maximum extent.  
In an SN regulated disk, $\cP\sim 1$;  a disk  with   $\cP\gg 1$  generates galactic fountains and winds and  thus reducing the amount of gas available for star formation, while $\cP \ll 1$ implies the disk gas is mostly  cold, 
 allowing the development of a surge of 
 star formation  until $\cP\sim 1 $ is reached. 
 This however refers to the case where the heating term in the energy balance equation is dominated by star formation feedback. Gravitational heating operating mainly at high $\Sg$,
 relaxes the condition of $\cP\sim 1$ and can allow  for a steady state with $\cP\ll 1$ by countering  dissipation.

The porosity  expression derived by  \cite{Silk1997} can be re-cast in the following form
\begin{equation}
\label{eq:SigmaSFRP}
\cP=(G \rhog)^{-1/2} \frac{\Sigma_{SFR}}{\Sg }\left(\frac{\sigma_f}{\sg}\right)^{2.7}\; ,
\end{equation}
where $\sigma_f$ is a fiducial velocity dispersion that is proportional to $m_{*}^{-1}$. 
For $m_{*} =250\,  \textrm{M}_\odot$, $\sigma_f=   22\kms$. Despite its importance,  estimation of porosity based on this expression is highly uncertain due to the steep  dependence on $\sg$ and $\sigma_f$. 
Taking $m_{*}=50\, \textrm{M}_\odot $ instead of $250\,  \textrm{M}_\odot$, reduces the porosity estimation by a factor  $\sim 80$ for the same $\Sfr$, $\sg$ and $\rhog$.

\begin{figure}
\includegraphics[width=.5\textwidth]{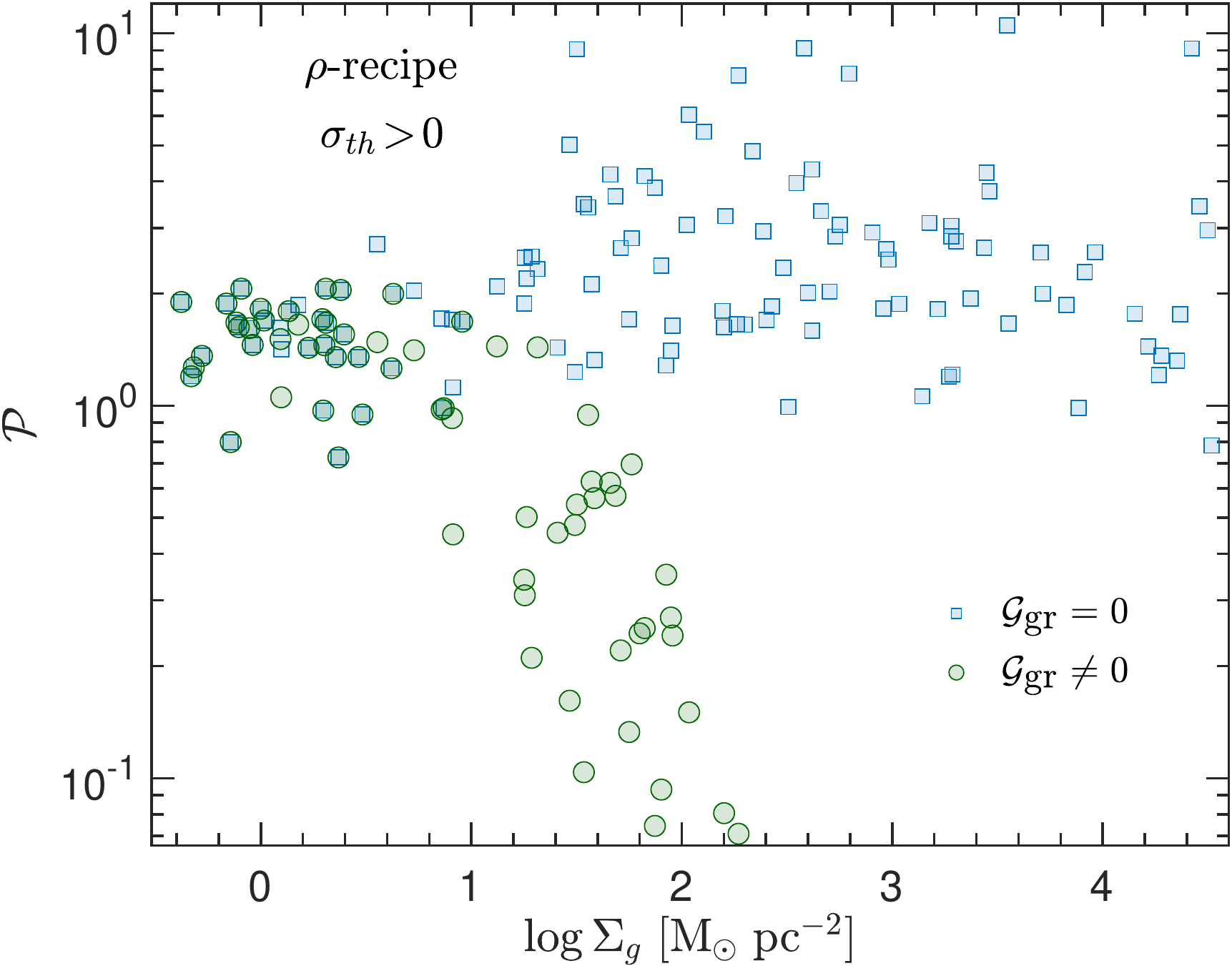}
\vskip 0.5cm
\caption{The porosity of the ISM versus $\Sg$ in the $\rho$-recipe obtained using the expression in Eq.~\ref{eq:SigmaSFRP} where the velocity dispersion, $\sg$, is 
obtained from the energy balance equation. 
Blue squares correspond to SN heating only ($\cGgr=0$), while 
the green circles are derived with gravitational heating switched on  ($\cGgr\ne 0$) in addition to SN feedback.
 }
 \label{fig:Po}
\end{figure}

Nonetheless, it is prudent to confirm that the porosity in our model is reasonable. 
In Fig.~\ref{fig:Po}, we plot the $\cP$ obtained by applying  the expression in Eq.~\ref{eq:SigmaSFRP} on our random sets of parameters where $\sg$ is, as before, computed from the energy balance equation with and without gravitational heating. 
The figure  refers to  $\sigth=5\kms$ for  the $\rho$-recipe only. For $\sigth=0$ the
porosity would be limited by the turbulence velocity dispersion $\sigt$  alone. 
According to Fig.~\ref{fig:sigv},  $\sigt\sim 3\kms$ at low $\Sg$, leading to very large $\cP$ if $\sigth=0$. The $Q$-recipe leads to similar results except that there are fewer points at low $\Sg$. The scatter of $\cP$ for the same $\Sg$ seen in the figure is almost entirely due to  $m_*$ which is distributed randomly in the range $70-200\, \textrm{M}_\odot$.  Smaller $\cP$ values are associated with larger $m_*$. 

For the $\cGgr=0$ case (blue squares), the porosity acquires reasonable values for all $\Sg$, with a declining trend toward higher $\Sg$.
Although  $\cP\propto \Sfr/\Sg \sim \Sg^{0.5}$ for $\cGgr=0$, the mild  increase in the velocity dispersion with $\Sg$ (see Fig.~\ref{fig:sigv}) is sufficient to counter the term $\Sfr/\Sg$.

The inclusion of gravitational heating has a dramatic effect on  $\cP$ at 
$\Sg\gsim \Sgth$ (green circles). 
The significant boost in 
$\sigt$ at  $\Sg\gsim \Sigunits$ due to {gravito-turbulence}  (Fig.~\ref{fig:sigv}),
and the steep dependence of $\cP$ on $\sg$, are the reason for the 
 sharp drop in $\cP$ at high $\Sg$. 

Porosity remains $\sim 1$ for the $\Sg$ range appropriate for normal spirals. 
According to Fig.~\ref{fig:sfrQ} starbursts with surface densities in the range  $\Sg \sim 100-300\Sigunits$ are curiously  matched by the model with $\cGgr=0$ (blue squares) while the model including gravitational heating (green circles) underpredicts $\Sfr$. Therefore, it is possible that 
SN-feedback remains dominant for starbursts in this range of $\Sg$.
If this is the case then we expect $\cP\sim 1$ in these starbursts. 
At higher densities, SN-feedback weakens relative to gravitational heating
and we expect $\cP\ll 1$ according to Fig.~\ref{fig:Po}.

\section{comparison with the other models in the literature}{
\cite{Forbes2014} and \cite{Krumholz2018} distinguished between star formation in the GMC and the Toomre  regimes. The time-scale in the former regime is fixed at a constant value of $2 \rm Gyr$, while the latter is determined  by the midplane gas density $\rho_g$ (which they express in terms of $Q$ and the angular velocity of the galaxy, among other parameters). 
The Toomre regime is adopted whenever the corresponding star formation time-scale if shorter than the fixed value of $2 \rm Gyr$.  The time-scale in their Toomre regime is actually the counterpart of our $\rho$-recipe. Our Fig.~\ref{fig:sigv}  implies that  $\rho_g \sim (\Sg/\sg)^2$ is an increasing function of $\Sg$. Thus   this regime is relevant to high $\Sg$ galaxies. According to   Eq.~50 in \cite{Krumholz2018},  the rate (per unit area) of gravitational heating 
via transport  is $\sim \sg^3 \Omega^2 $ in the limit of high $\Sg$,
which is proportional to the heating rate implied in the $\rho-$recipe. Therefore, in this limit of $\Sg$, we expect our model to yield similar scaling as the transport model by \cite{Krumholz2018}. Indeed, the  slopes of  $\log \Sfr$  versus $\log \Sg$ are similar as revealed by comparing Fig.~\ref{fig:sfrQ} here with Figs.~2\&3 in \cite{Krumholz2018}. Further, the velocity dispersion derived in the two models are 
in the same range. Another similarity between the two models is that 
SN heating dominates at low $\Sg $ below a  threshold value ($\Sgth$ in our model). In this limit, the dependence of $\Sfr$ on $\Sg$ approaches $\Sg^2$ and the velocity dispersion acquires $\sg\sim 5-10 \kms$ with weak dependence on $\Sg$. This scaling can readily by derived by equating 
the weight of the disk by the vertical momentum injection by SN in the midplane \citep{Ostriker2011}. However, we emphasize  that even in the SN dominant limit, vertical balance alone is insufficient to yield $\sg$ to close a self-regulating star formation versus  feedback loop. }

\cite{Bacchini2020} present a detailed analysis of 
the kinematic of ten nearby star forming galaxies. They find that SN feedback is sufficient to maintain the observed turbulence level. The gas surface density in these galaxies is lower than the threshold value obtained here, therefore, these findings  are actually consistent with our model. However, we note that, in contrast us,  their 
model of star formation  induced turbulence is based on 
SN energy output rather than momentum injection into the ISM. 

{Star formation models based on the turbulence properties of MCs have been developed \citep[e.g][]{Krumholz2005,Padoan2011,Federrath2012,Padoan2017,Burkhart2018}. These models rely on an assumed (but simulation-motivated) form for the  probability distribution of the gas density inside
 turbulent MCs, with star formation ensuing in sub-cloud regions with 
 sufficiently high densities. These considerations can be 
 used to provide the global star formation rate in the disk under certain  conditions  regarding  marginally disk stability and the fraction of molecular gas.
 Turbulence within individual MCs plays an important role in their  {overall } stability  and establishing the distribution of density fluctuations
inside them
 \citep{Zuckerman1974,Fleck1980, Scalo2004,Larson1981,Bonazzola1987,Falgarone1990,Krumholz2005,Hennebelle2011, Federrath2012,Mocz2017,Padoan2017}. 
{This} turbulence can be triggered by global gravitational and MHD instability  over large scales of 100s of parsecs. 
Further, at  high SN rate and high gas surface densities, turbulence can be boosted by 
 nearby SNs \citep[e.g][]{Ballesteros-Paredes2007,Seifried2018}. 
 Simulations demonstrate that {gravito-turbulence}  within  MCs
 is necessary to explain the Larson relations \citep{Larson1981} between the observed internal velocity dispersion and sizes  \citep{Ibanez-Mejia2016}.
 }
 
{In the limit of SN dominated heating, porosity of SN remnants must be $\sim 1$ if self-regulated feedback is maintained \cite{Silk1997}. Together, with a requirement of a  marginally stable disk for maintaining star formation, it is possible to derive closed forms for the velocity dispersion and star formation law. Our model here is in agreement with this picture at $\Sg\lsim \Sgth$, but deviates from it when gravitational heating is included.}
 
\section{Discussion} 

In this paper,  we have  highlighted  the role of self-gravity induced turbulence as a mechanism for regulating star formation in order to bring down the SFR relative to  models invoking only SN feedback.
However, it is known that turbulence can have  the opposite effect of triggering star formation  \citep{Elmegreen2002,Federrath2010}. Convergent turbulent motions  in the ISM 
lead to large scale agglomeration of dense and cool gas, facilitating the   conditions for star formation. 
 AGN outflows also provide  potential triggers of gas-rich disk star formation when the interstellar medium pressure  is transiently enhanced \citep{Silk2009,Ishibashi2013, Dugan2014,Bieri2016, Zubovas2017}. We will address the possible role of AGN elsewhere.   

Here we note that  that reconciliation of gravity-driven turbulence  with the possible quenching or   triggering of star formation depends on the 
detailed statistical properties of the density and velocity
distribution in galactic 
turbulence. As pointed out by \cite{Ostriker2011}, moderate densities generated by  converging flows are likely to be dispersed before they can  collapse to form stars. Thus, only  a small fraction of the gas  in the densest regions would be available for star formation via this mode. 
We note here that this actually might help explain the low star formation efficiency on galactic scales using the arguments of  \cite{Krumholz2005}, but for turbulence on large scales rather than in MCs.

Turbulent motions driven by Toomre   gravitational instabilities occur on a relatively large scale. For a Toomre parameter $Q\lsim 1$,
 the disk scale-height is close to the  driving scale, 
and turbulence should 
be  three-dimensional and  dissipate  on  small scales 
through   a downward  cascade of energy
  \citep{Elmegreen2003}. We have  not considered the possibility of a quasi-two-dimensional inverse  cascade turbulence initiated  on  scales
  larger than the disk height. A quasi-two-dimensional turbulence could be driven by long spiral arms \citep{Bournaud2010} and even become dominant if $Q>1$.

{\mbox{\cite{Brucy2020}} use numerical simulations to study the suppression of star formation by gravito-turbulence. Their study illustrates the  insufficiency of SN feedback alone to regulate star formation at high surface densities. In their simulations (of  a cube of  $1\kpc$ on the side) gravito-turbulence  driving is introduced  by numerically injecting energy in the system rather than self-consistently as the result of gravitational instability. 
They consider  a disk of marginal instability with $Q\sim 1$, 
 in contrast to our model where the Toomre parameter can actually be less than unity at large $\Sg$.}

We have focused on a possible route for SFR regulation based on the dynamical properties of disks on large scales. 
Nonetheless, there are several layers to the process of star formation, involving a wide range of physical properties. 
Star formation occurs exclusively in MCs with  sizes of tens of parsecs and mean number densities of 
$\mathcal{O}(100)\;  \textrm{cm}^{-3}$, but with large variations of up to $10^5 \;  \textrm{cm}^{-3}$. Inside MCs, stars preferentially form in the densest regions 
over physical scales below  hundreds of   au,   where the details of the local equation of state, metallicity and photoionizing feedback  play a fundamental role in determining the 
initial stellar mass function
\citep[e.g.][]{Dale2005,Hennebelle2019}. Furthermore, there is an intricate interface between the MCs and the interstellar (ISM) medium on large scales. Radiative, gravitational and MHD processes on large scales likely dictate the formation of the MCs and their subsequent properties \citep{Ballesteros-Paredes2007}.

The standard disk gravitational  stability analysis assumes a fixed $\sg$ in the 
perturbed equations without perturbing $\sg$ itself. 
In principle, with the restrictions imposed by  Eq.~\ref{eq:GainLoss}, a 
 a full stability analysis should incorporate   perturbations in  $\sg$. This will introduce additional (thermal) instabilities  on small scales and also modify the 
stability criteria on intermediate scales \citep{Nusser2005b}. However, this is a lengthy analysis  that is not expected to have a significant effect on the results.

Galactic outflows   launched by the cumulative effect of stellar and/or AGN feedback are ubiquitously observed \cite[c.f.][for a review]{Rupke2018}. They likely have a 
profound impact on the global properties of galaxies, such as the their baryonic content, the DM density  distribution and 
the overall star formation history. The model presented here assumes that the disk  relaxes to a  quasi-stationary state dictated by vertical pressure and energy balance.
This quasi-equilibrium state will be transiently disturbed by 
rapid episodes of galactic winds,  but it is assumed to persist on in the  long time average sense. This assumption is partially supported by numerical simulations \citep[e.g][]{Smith2018}.

One of the interesting observations in star formation is the long depletion time for turning gas into stars. 
The depletion time-scale is $\sim 20-100 $ times larger than any dynamical timescale in the system, i.e. the star formation efficiency or the fraction of gas that actually turn into stars per dynamical timescale is very small. 
We do not make an attempt at explaining this issue and rather represent  the star formation efficiency as a tunable parameter.  
The explanation for the low efficiency is most likely related to the susceptibility of MCs  to stellar feedback (winds and photoionization) even before massive stars explode into supernovae  \citep{Elmegreen2000,Inutsuka2015,Kruijssen2019,Chevance2020}.
Some compelling evidence for this is presented  by    \cite{Kruijssen2019}  for the spiral galaxy NGC 300 \citep[but, see also][]{Koda2009}.
Nonetheless, we have found (see Eq.~\ref{eq:noeps}) that at low disk surface densities and 
for a high gas fraction, 
the SFR per unit area is independent of the efficiency parameter and is dictated by 
the parameter $\vsn$ related to the momentum injected by SN into the ISM.  The SFR in this case is also reasonable, and hence  under certain conditions, the long depletion time-scales could be  related to large-scale feedback in the disk rather than within MCs.

According to the SK law, a typical galaxy depletes  its gas content over  a time-scale of $\Sg/\Sfr\sim 6 \Sg^{-0.41}$ Gyr ($\Sg$  in $M_\odot\textrm{pc}^{-2}$) \cite{Schmidt2016}.
Therefore, at least for high $\Sg$, refilling of the disk gas content is inevitable. 
Since on average, all galaxies will consume their gas, it is unlikely that galaxy mergers
can provide the required gas supply. 
This motivates  direct accretion of gas from the intergalactic medium as the main replenishment  mechanism \citep{Bournaud2009,Dekel2009,Conselice2012,Lofthouse2017}.
 Feeding of warm galactic coronae is an intermediate phase with potentially observable consequences. 
The circumgalactic medium may indeed control the galactic baryon budget \citep{Tumlinson2017}. 
Transport of accreted gas to the central disk has been observed in some galaxies 
\citep{Schmidt2016}. Further, cosmological zoom-in simulations of Milky Way-like galaxies find that gas accretion replenishes the supply of gas for disk star formation. The angular momentum of the accreted gas is similar to that at the edge of the gaseous disk  \citep{Trapp2020}.
The model of \cite{Krumholz2018} builds on the gravitational energy released via the transport in a sheared disk as the main source of {gravito-turbulence}.

A robust prediction of the model is that disks of normal spirals above a threshold surface density
 $\Sg\sim 50\Sigunits$, cannot  be regulated by SN feedback if the observed SK law is to be recovered. {Gravito-turbulence}  dominates these disks, leading to $\Sfr$ consistent with the observations.
Below this threshold, the disks are marginally stable with $Q\approx 1$. 
At higher surface densities, only disks dominated by the stellar component
have $Q\approx 1$. Disks above the threshold  but having a low ratio $\Sast/\Sg$, are associated with $Q<1$.

The situation is  different for starburst galaxies. 
Starbursts with $\Sg \sim 100-300\Sigunits$ are associated with $\Sfr$ predicted by SN feedback alone. In fact, 
gravitational heating suppresses the predicted $\Sfr$ by about an order of magnitude relative to the observations of starbursts in this range of surface densities.  However, at larger $\Sg$, starbursts tend to fall back to the $\Sfr$ level predicted by gravitational heating and consistent with an extrapolation of the SK law inferred for normal galaxies ($\Sg\lsim 170 \Sigunits$)

The results are only moderately  sensitive to the dissipation parameter $\etat$. An application of  the model on the random sets of disk parameters  with $\etat=3.5$ instead of the default  $\etat=\etavQ$ leads to meagre  differences in the results (see also Fig.~\ref{fig:epsdep}). Although, the sensitivity to $\epsf$ is more pronounced over the range of surface densities we consider, there is no need for fine-tuning. Variations of  $\lsim \, 20\%$ in $\epsf$  yield reasonable matches to the star formation law, even with    $\etat$ fixed to the default value. 

\section{acknowledgements}
This research is supported by a grant from the Israeli Science Foundation grant number 936/18.

\section{Data availability}
No new data were generated or analysed in support of this research.

\bibliographystyle{mnras}

\bibliography{references.bib}

\appendix
\section{ approximate expressions, scale of least stable mode and disk height}
\label{sec:tests}
\begin{figure}
\includegraphics[width=.5\textwidth]{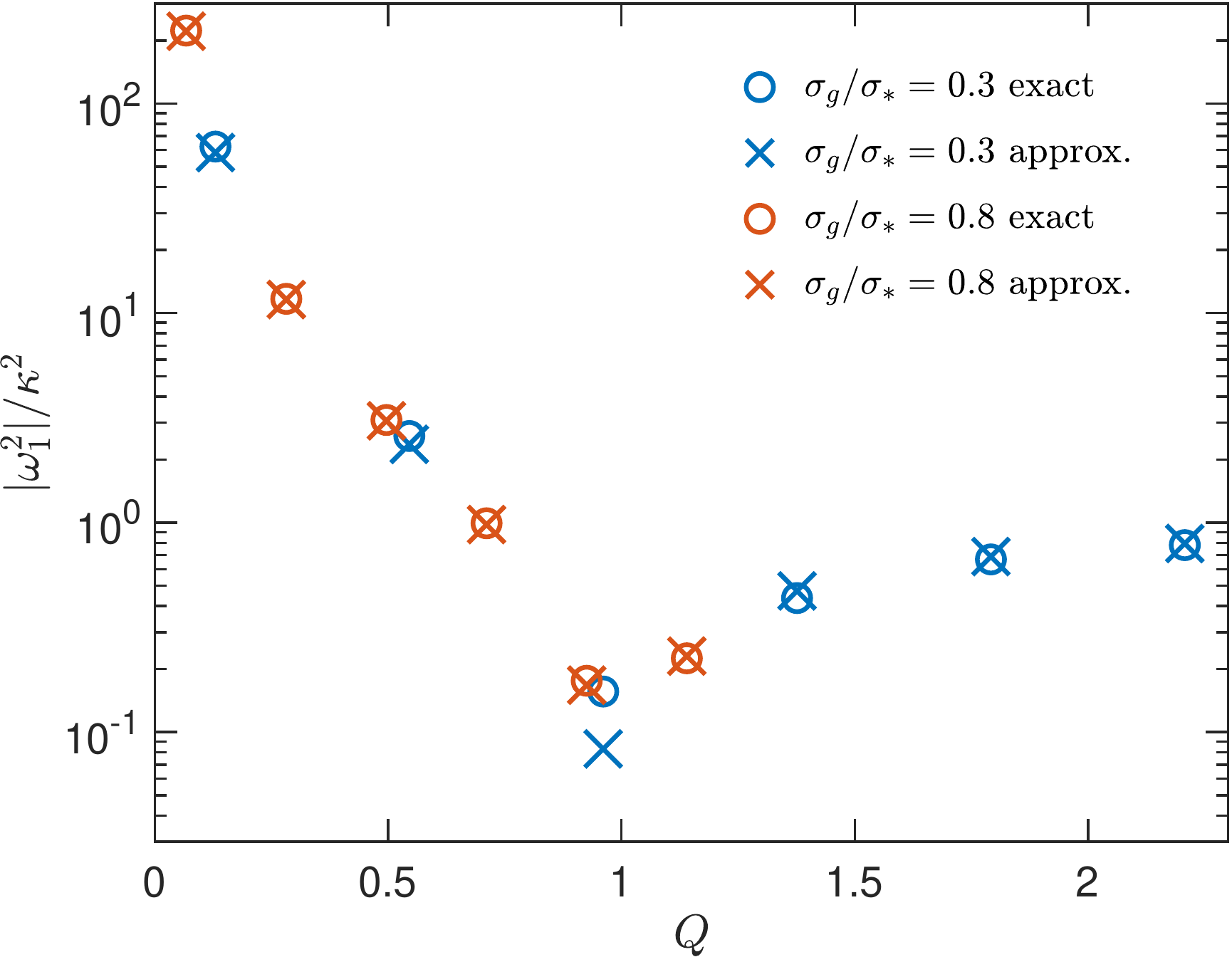} 
\vskip -0.8cm
\includegraphics[width=.5\textwidth]{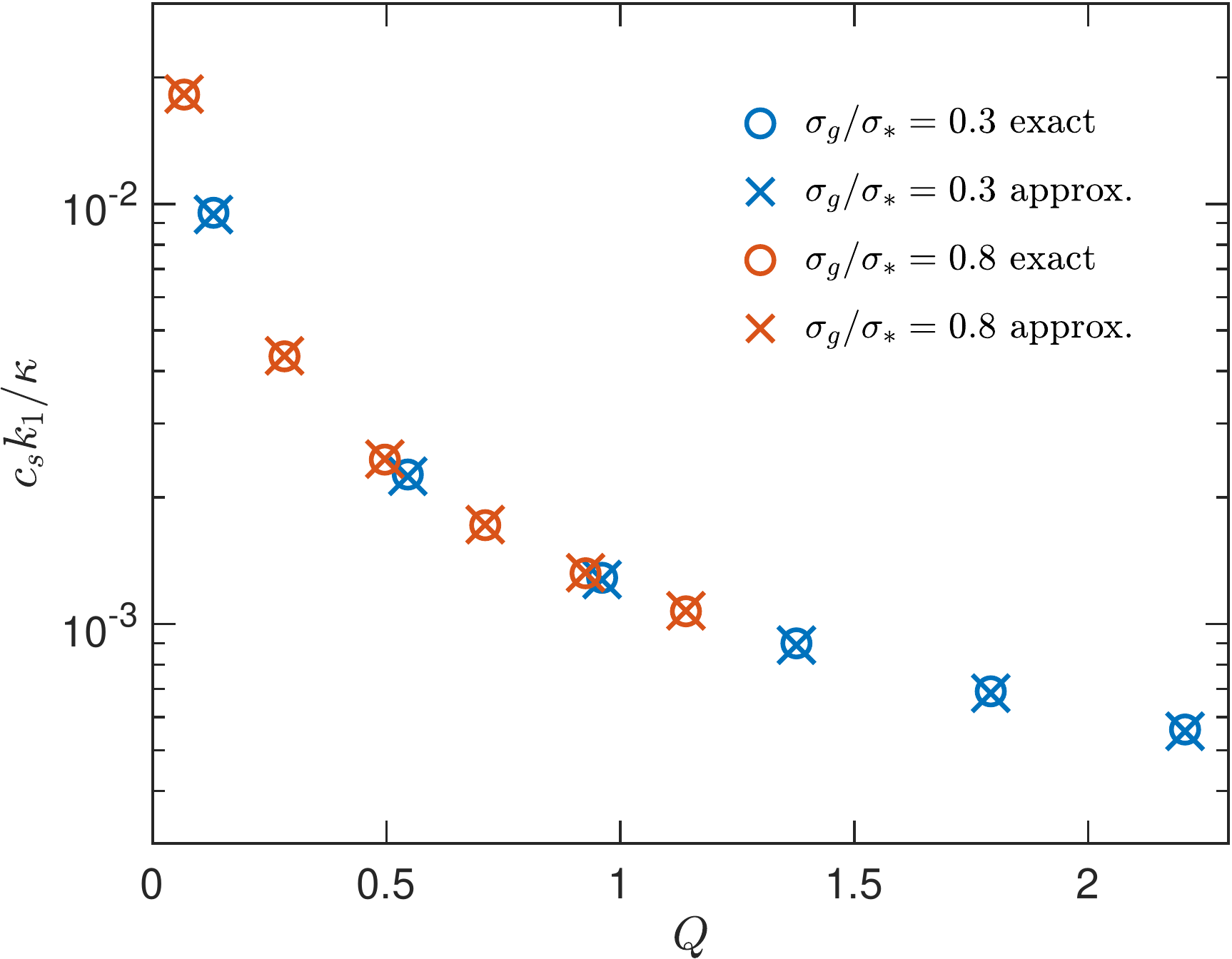}
\caption{Top: accuracy of the approximate expression in Eq.~\ref{eq:tgrow} for  $\omega_1$ 
as a function of the Toomre parameter $Q$ in a two fluid  system with fixed $\kappa$, $\Sast$ and $\Sg$ (see text), for two values of  $\sg/\sast$, as 
indicated in the figure. 
Bottom: test of  the expression in Eq.~\ref{eq:kone} for the wavenumber $k_1$  of the least stable mode.}
 \label{fig:omegatest}
\end{figure}

We present tests of  the approximate relations in Eqs.~\ref{eq:ceff}--\ref{eq:kone} for gravitational instability in a composite two-fluid disk.
Fig.~\ref{fig:omegatest} presents such a comparison for a composite disk with  $\kappa = 0.035 \kms \, \textrm{pc}^{-1}$, 
$\Sg=15\Sigunits$ and $\Sast=30 \Sigunits$.
 The quantities $\omega_1$, $k_1$ and $Q$ are  computed for a range of $\sg$ for two fixed values of the ratio $\sg/\sast$. 
In the top panel, we plot  $|\omega_1^2|/\kappa^2$ versus $Q$ 
where the 
crosses and open circles, respectively,  correspond to  the approximation in Eq.~\ref{eq:tgrow} and the exact result obtained from the exact relations in  \cite{Jog84}. Only  at $Q\approx 1$ for $\sg/\sast=0.3$, is there  a significant difference between the approximate and exact results, but this is  irrelevant in practice  since $\omega_1\approx 0$ at this $Q$. Since the dependence on $\kappa$  and $Q$ in Eq.~\ref{eq:tgrow} is exact in linear theory, the plot essentially tests the validity of the \cite{Romeo2011} approximation in  Eq.~\ref{eq:Qeff} for $Q$. Although  this approximation has been tested previously by these authors, we bring this comparison here simply to emphasize that it is also suitable for estimating  $\omega_1$.
The bottom panel of the same figure, shows  $k_1$ in units of $\kappa/c_s$. Here also, the  agreement between the approximate expression in Eq.~\ref{eq:kone} and the exact result based on \cite{Jog84} is excellent.

\begin{figure}
\includegraphics[width=.5\textwidth]{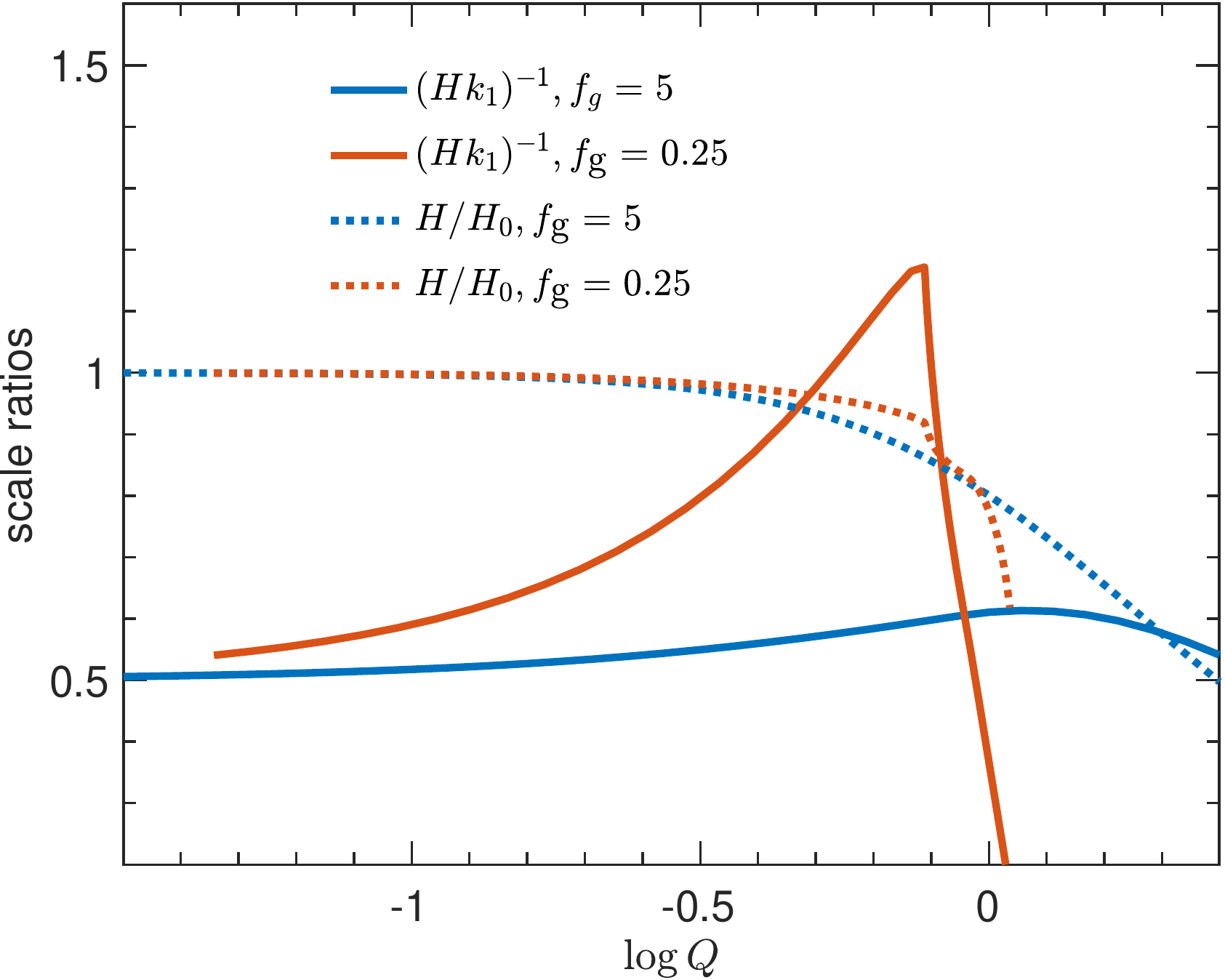} 
\vskip 0.5cm
\caption{The solid curves show the  ratio of  $1/k_1$  
 to the disk scale height (including halo gravity), $H$. 
The dotted curves are  $H/H_0$ where $ H_0$ is  computed for zero  halo gravity. Results are plotted for two values of $f_g$, as indicated in the figure. }
 \label{fig:Hrat}
\end{figure}

In Fig.~\ref{fig:Hrat} we plot the scale ratios $H/H_0$ (Eq.~\ref{eq:dHwh}) and $(H k_1)^{-1}$ versus the Toomre parameter, $Q$. 
The curves correspond to two  values of $f_\textrm{g}=\Sg/\Sast$  as indicated in the figure. For each $f_\textrm{g}$,
 the ratio of the velocity dispersion  $\sg/\sast$  is varied to produce the corresponding curve. 
 We choose $\sast$ such that $ \sg/\sast=1$ yields $Q=1$ and hence all points with $Q>1$ ($Q<1$)  have $ \sg/\sast>1$ ($\sg/\sast<1$). As for the DM halo, we take $v_\textrm{h}=100\kms$ at $R=3\; \textrm{kpc}$. 
 Deviation of $H/H_0$ from unity is an indication of the halo gravity on the disk height. The figure demonstrates that for unstable disks ($Q\lsim 1$), halo gravity has a minor effect on the disk height.
Further, we see that  $1/k_1\sim H$ as long as $ Q\lsim 1$ and even larger $Q$ for a gas dominated disk as indicated by the solid red curve corresponding to $f_g=5$.

\section{depletion time, gain and loss terms}
\label{sec:depG}

\begin{figure}
\includegraphics[width=.5\textwidth]{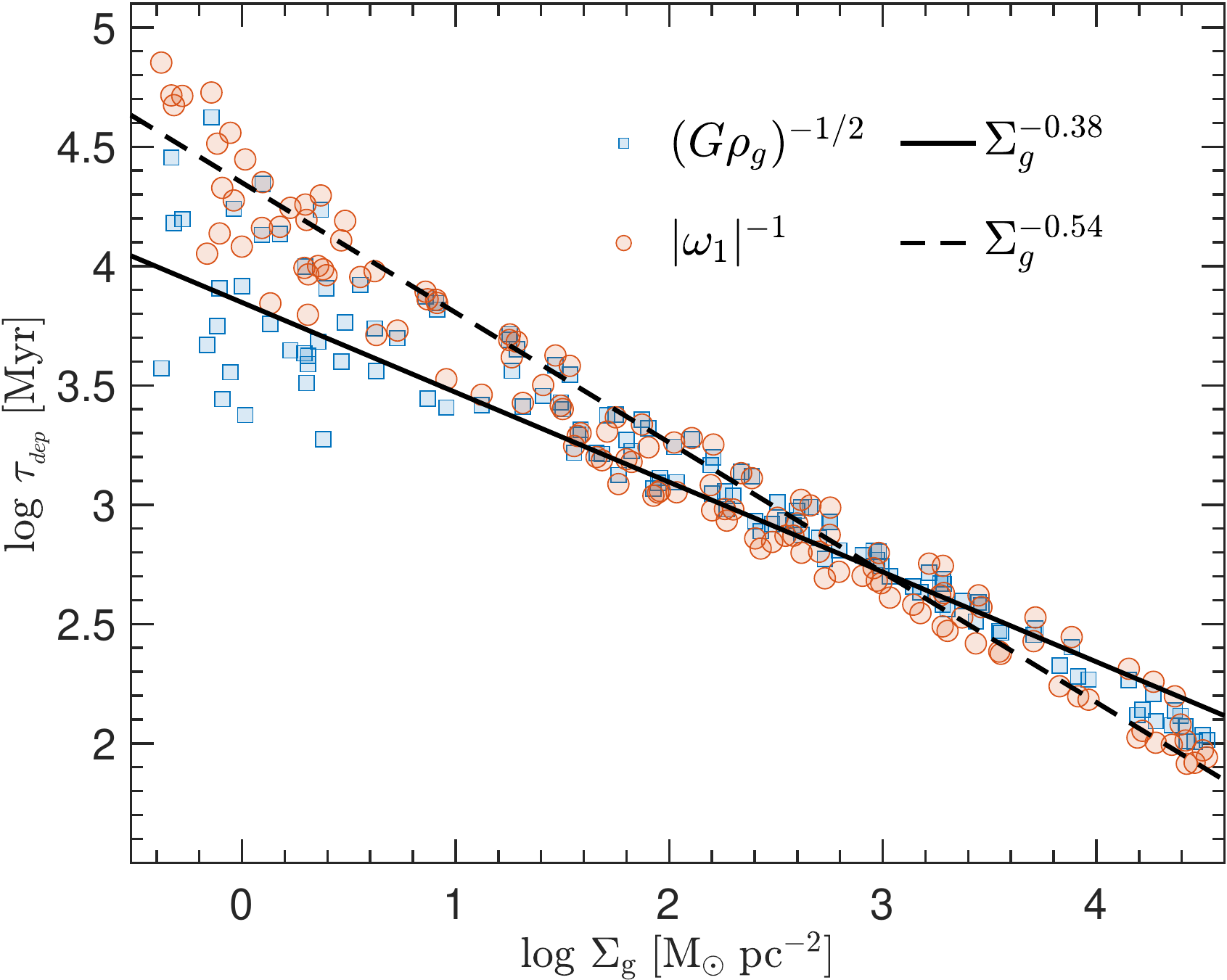} 
\vskip 0.5cm
\caption{The gas depletion timescale $\tau_{dep}=\Sg/\Sfr=\tsf/\epsf$ for $\sigth=0$.}
 \label{fig:tSF}
\end{figure}

Fig.~\ref{fig:tSF} examines in more details  the differences in $Q$- and $\rho$-recipes by plotted the  depletion timescale $\tau_{dep}=\tsf/\epsf$ for points shown in the previous figure. The main difference is at the low $\Sg$ where the ratio is on average lower in the $\rho$-recipe. At larger $\Sg$, the two recipes are essentially equivalent.  This is in accordance with Eq.~\ref{eq:Hkratio} and Fig.~\ref{fig:Hrat} showing that $k_1\sim H^{-1}$.

\cite{Leroy2017} and \cite{Querejeta2019} analyzed the  star formation in the spiral galaxy M51. They find a mild anti-correlation between the molecular gas depletion time and the molecular gas surface density 
on the cloud scale. Our depletion times are  in the same ballpark for the same range of 
$\Sg$ in Fig.~\ref{fig:tSF}, but we find a stronger anti-correlation. 
Although  most of the gas mass within $10 \kpc$ in M51 is molecular \citep{Scoville1983}, the surface density in our work 
is to be  understood as an average over larger scales, rather than the individual molecular cloud scale. The reason is that disk stability addressed here depends on the average 
surface density rather than local densities of the clouds. 

\begin{figure}
\includegraphics[width=.5\textwidth]{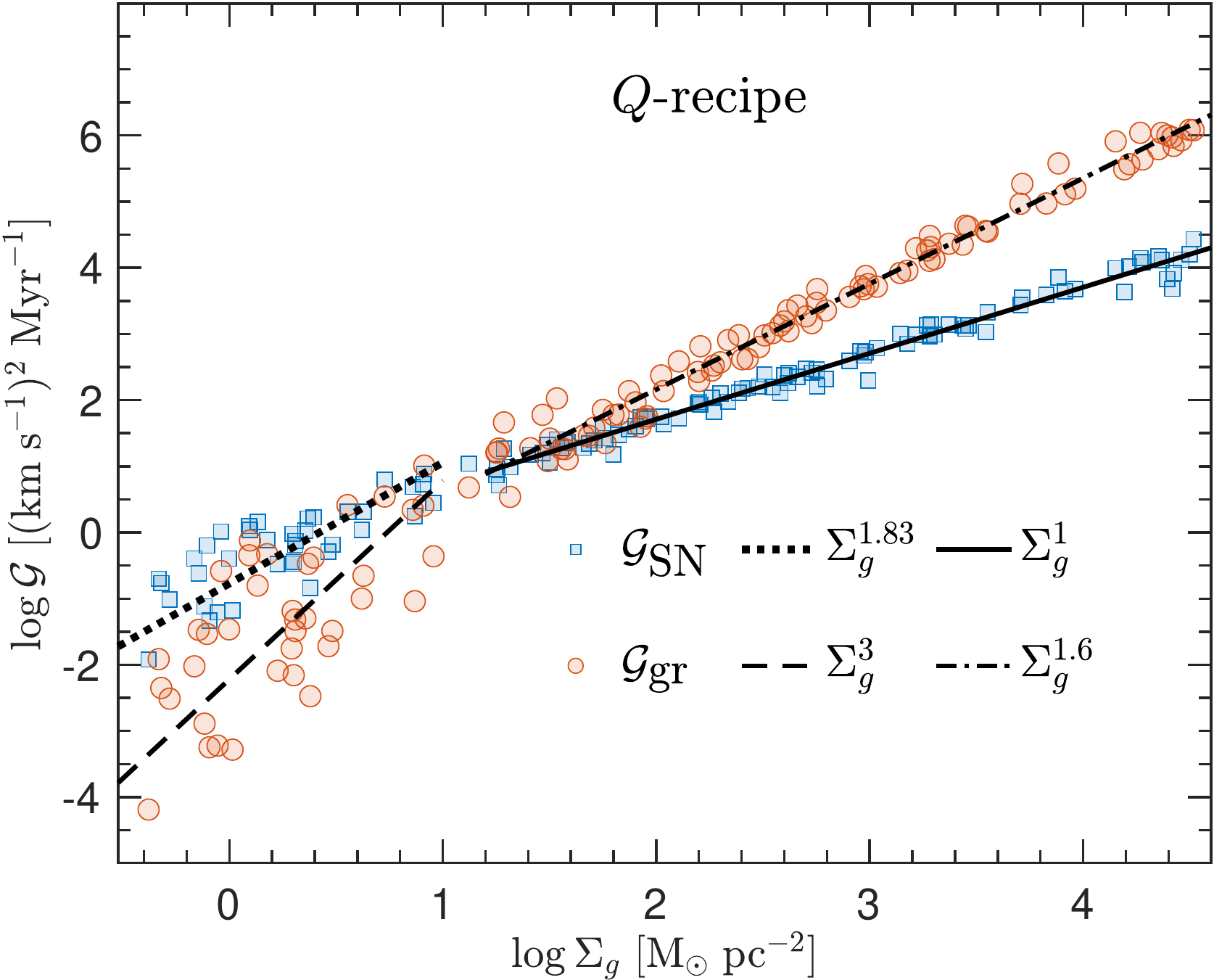} 
\vskip 0.5cm
\caption{The energy injection rates (per unit mass) due to SN and gravitational heating, as indicated in the legend. $\sigth=0$}
 \label{fig:G}
\end{figure}

We turn now to a direct comparison between  the SN and gravitational heating terms, $\cGsn$ and 
$\cGgr$. Qualitatively, the picture is similar in both recipes and thus we present results only for the $Q$-recipe. 
The blue squares and  amber circles in Fig.~\ref{fig:G} represent, respectively,  $\cGsn$ and $\cGgr$, obtained with $\sigth=0$ in the $Q$-recipe with the default parameter values. 
The points correspond to  the top panel in Fig.~\ref{fig:sfrQ}.
As inferred previously from Fig.~\ref{fig:sfrQ},   $\cGgr$ becomes more significant only 
at $\Sg>10\,  \Sigunits$. 

Eq.~\ref{eq:Qlimit} allows us to derive the dependence of the heating terms at low surface densities. 
Since $\cGgr= \omega_1 \sigQ^2$, the equation implies $\cGgr\sim  \Sg^2 (Q^{-2}-1)^{1/2}\kappa^{-1}\sim \Sg^3$ if the disk is 
purely gaseous and the contribution from $\Sg$ to $\kappa$ is negligible. This is steeper than the slope of the linear regression  line (dashed)
of $\log \cGgr$ on $\Sg$, obtained  for the amber circles  restricted to $\log \Sg <1$. 
This is reasonable given our assumptions and the large scatter of the amber circles representing $\cGgr$ in the figure. 
We have checked  that in fact  linear regression yields a slope very close to $3 $ when  limited  to points in a narrow range of  $\kappa$  and $\fg$. 
Similar considerations imply $\cGsn \sim  \Sg^2$ which is steeper than the slope of the dotted line but, as before, this is due to the spread in 
$\kappa$ and $\fg$. 

At large $\Sg$, the epicyclic frequency is mainly fixed by $\Sg$, i.e. $\kappa \sim \Sg^{-1/2}$. Considering only the gravitational heating term in a gas-only disk, energy balance yields  a Toomre parameter which depends just on $\etat$, i.e. $Q=Q(\etat )$ and hence
$\cGgr\sim \kappa Q(\etat)\Sg^2\sim \Sg^{1.5}$, close to the slope of the dot-dashed line in Fig.~\ref{fig:G}.
Further, $\sg\sim Q(\etat)/\kappa \Sg$ and hence  $\cGsn\sim \kappa [Q(\etat)^{-2}-1)^{-1/2} \sg\sim \Sg$ in excellent agreement with the slope of the solid line.

\bsp	
\label{lastpage}
\end{document}